\documentclass[preprint,superscriptaddress]{revtex4}
\usepackage{graphicx}

\begin{document}

\title{Steady states and oscillation modes of two driven dissipative oscillators with non-Hermitian coupling}

\author{Sahel Ashhab}
\affiliation{Advanced ICT Research Institute, National Institute of Information and Communications Technology, 4-2-1 Nukui-Kitamachi, Koganei, Tokyo 184-8795, Japan}
\affiliation{Research Institute for Science and Technology, Tokyo University of Science, 1-3 Kagurazaka, Shinjuku-ku, Tokyo 162-8601, Japan}

\author{Aakanksha Sud}
\affiliation{Frontier Research Institute for Interdisciplinary Sciences, Tohoku University, 6-3 Aoba, Sendai 980-8578, Japan}
\affiliation{Research Institute of Electrical Communication, Tohoku University, 2-1-1 Katahira, Sendai 980-8577, Japan}

\author{Shunsuke Fukami}
\affiliation{Research Institute of Electrical Communication, Tohoku University, 2-1-1 Katahira, Sendai 980-8577, Japan}
\affiliation{WPI-Advanced Institute for Materials Research, Tohoku University, Sendai 980-8577, Japan}
\affiliation{Center for Science and Innovation in Spintronics, Tohoku University, Sendai 980-8577, Japan}
\affiliation{Center for Innovative Integrated Electronic Systems, Tohoku University, Sendai 980-8572, Japan}
\affiliation{Inamori Research Institute for Science, Kyoto 600-8411, Japan}

\date{\today}

\begin{abstract}
We analyze the dynamics of two harmonic oscillators with nonlinear, non-Hermitian coupling between them. Specifically, we consider a synthetic antiferromagnet composed of two ferromagnetic layers with antiferromagnetic interactions between the two layers, leading to the emergence of two oscillation modes of the combined system. We analyze the response of this system to external driving fields. We determine the allowed steady states under various driving conditions. We then analyze the dynamics of small deviations away from the different steady states, which reveals the normal modes that can be probed using spectroscopic techniques in experiment. As expected, the nonlinear system exhibits multistability, in which multiple steady states can exist for the same driving conditions. We find that the dynamical response can be drastically different depending on which one of the two modes is driven and on the strength of the driving. Specifically, we can obtain level-repulsion or level-attraction patterns in the spectrum. In addition to the rich variety of spectra, we find that some steady states exhibit normal modes that contain multiple frequencies. Our results explain recent experiments on synthetic antiferromagnets and provide guidance for designing experiments that can explore previously unseen phenomena in these systems.
\end{abstract}

\maketitle

\section{Introduction}

Nonlinear dynamical systems exhibit a rich variety of motional patterns \cite{Strogatz,Alligood}. These include chaotic motion, where a small perturbation can lead to exponentially growing changes in the dynamics, bistability and multi-stability, where multiple stable steady-state dynamical trajectories exist, as well as frequency conversion, where a force of a certain frequency induces a response at a different frequency. One example of the latter is harmonic and sub-harmonic generation \cite{Boyd}. Another nonlinear dynamical effect is the synchronization of coupled oscillators \cite{StrogatzSynchronization}. In continuous media, nonlinear dynamics leads to phenomena such as turbulence \cite{Davidson} and stable non-equilibrium field structures, such as solitons \cite{Ablowitz} and skyrmions \cite{Mishra}. Nonlinear dynamical phenomena arise in quantum systems as well \cite{Drummond}. One of the prominent phenomena in this context is squeezing \cite{Clerk,Ashhab2025}, which has applications in quantum-enhanced sensing \cite{Jia}.

Nonlinear dynamics has been realized in a vast range of physical systems. A few recent examples include attosecond atomic physics \cite{Paul}, cold-atom Bose-Einstein condensates \cite{Khaykovich,Strecker} and fiber lasers \cite{Song}. The nonlinear behaviour of Josephson junctions is a crucial property that enables the use of these devices to construct qubits for quantum computing \cite{Krantz}.

As mentioned above, some of the dynamical phenomena in nonlinear systems can be utilized for practical applications, such as sensing. For example, by driving the system close to conditions of instability, the driven system can be highly sensitive to small signals that either keep it in its initial oscillation state or cause it to switch to a drastically different oscillation state. This mechanism has been used to detect the weak magnetic field signal produced by a superconducting qubit and distinguish between the two states of the qubit via the switching dynamics of a coupled nonlinear oscillator \cite{Lupascu,Lin}.

Another important phenomenon in the context of sensing that has been the subject of increasing research activity is the high sensitivity to parameter variations at bifurcation points, including exceptional points. The high sensitivity to external parameters can be understood intuitively based on the fact that, at bifurcation points, the characteristic frequencies often follow a square-root function of some system parameters, and the square-root function has an infinite slope at the bifurcation point. In addition to their potential practical applications, exceptional points have been the subject of increasing research activity recently in relation to the theoretical understanding of non-Hermitian physics \cite{Ashida}.

The present work is motivated by two recent experiments on driven synthetic antiferromagnets \cite{Sud2025,Sud2026}. While these two experiments were generally similar, they produced qualitatively different types of responses. In particular, one experiment, in which the two-mode system was driven near its acoustic mode frequency, exhibited a level-repulsion pattern in its spectral response \cite{Sud2025}. The other experiment, in which the system was driven near its optical mode frequency and its two-mode coupling was set in a different regime, exhibited a level-attraction pattern in its spectral response \cite{Sud2026}. This situation creates a need to perform a theoretical investigation to help understand the mechanisms and/or parameter regimes that give rise to level-attraction and level-repulsion patterns.

We show in this work that, as a result of the asymmetric form of the coupling mechanism, driving the two modes gives rise to two different types of response. Our theoretical results therefore explain the experimental features observed in Refs.~\cite{Sud2025,Sud2026}. Furthermore, we find even more exotic spectra that can be obtained under different driving conditions, for example strong driving that is currently out of reach for the experimental setup of Refs.~\cite{Sud2025,Sud2026}. We also find situations in which multi-frequency normal modes emerge in the dynamics of small perturbation around the steady state. These phenomena could be observed in future experiments.

The remainder of this paper is organized as follows: In Sec.~\ref{Sec:Model} we describe the theoretical model and equations of motion for the dynamical system. In Sec.~\ref{Sec:Optical}, we consider the case when the driving field is applied to the optical mode. We analyze the steady states, the dynamics of perturbations around the steady states and the resulting normal mode spectra. In Sec.~\ref{Sec:Acoustic}, we perform similar analysis for the case when the driving field is applied to the acoustic mode. We give some concluding remarks in Sec.~\ref{Sec:Conclusion}. In the appendices, we treat two simpler models to elucidate which phenomena are associated with which elements in various theoretical models.

\section{Mathematical model of the dynamical system}
\label{Sec:Model}

\begin{figure}[h]
\includegraphics[width=14cm]{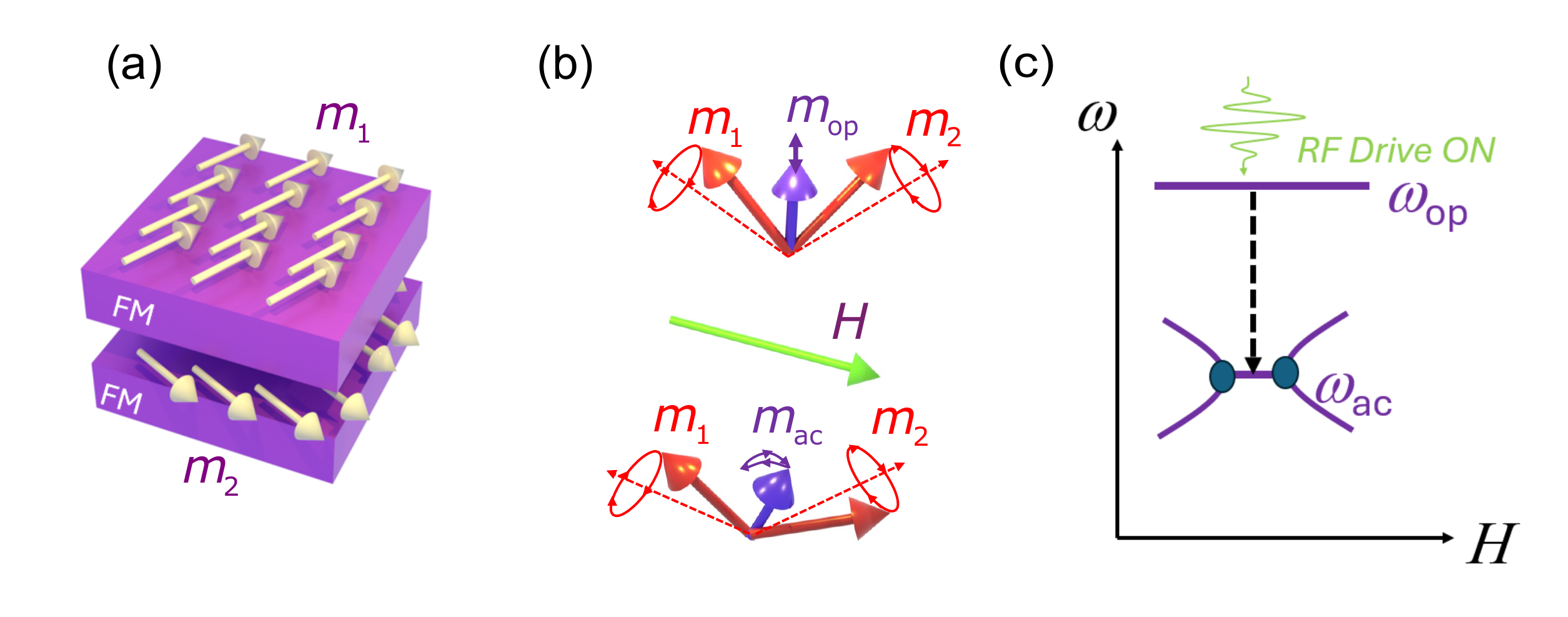}
\caption{Schematic diagram of synthetic antiferromagnet. (a) Two ferromagnetic (FM) layers (with magnetizations $m_1$ and $m_2$) are coupled antiferromagnetically. (b) The two coupled magnetizations form two collective oscillation modes, namely the acoustic (ac) and optical (op) modes. An externally applied magnetic field ($H$) can be used to tilt the two layer magnetizations and, as a result, modify the frequencies of the two modes ($\omega_{ac}$ and $\omega_{op}$). (c) A radio-frequency (RF) magnetic field is used to drive the system. By adjusting the static and driving fields, a variety of dynamical responses can be obtained. One example shown here is the level attraction pattern in the oscillation frequency spectrum.}
\label{Fig:SchematicDiagram}
\end{figure}

We consider a synthetic antiferromagnet composed of two ferromagnetic layers that interact antiferromagnetically \cite{Chiba,Gallardo,Shiota,Sud2020,Fukami,Dai,Li,Awschalom,Sud2023,Millo,Sui,Wang,Devolder}. The system is illustrated in Fig.~\ref{Fig:SchematicDiagram}. Each layer has a collective magnetization. The interaction between the two layers gives rise to two hybridized oscillation modes: a low-frequency mode commonly called the acoustic mode and a high-frequency mode commonly called the optical mode. The equations of motion for these normal modes contain the usual harmonic-oscillator, driving and dissipation terms. Owing to the nonlinear nature of the interactions between the two magnetization vectors of the two magnetic layers, the equations of motion for the two normal modes contain nonlinear inter-mode coupling terms.

Following Ref.~\cite{Sud2025}, our starting point for the theoretical modeling of the dynamics in this work are the two coupled equations of motion for two driven dissipative nonlinearly coupled oscillators, i.e.~Eqs.(1,2) of Ref.~\cite{Sud2025}:
\begin{eqnarray}
i \frac{db_{ac}}{dt} & = & \left( \omega_{ac} - i \kappa_{ac} \right) b_{ac} + 2 i g_3 b_{ac}^* b_{op} + \tau_{ac},
\label{Eq:EOMac}
\\
i \frac{db_{op}}{dt} & = & \left( \omega_{op} - i \kappa_{op} \right) b_{op} - i g_3 b_{ac}^2 + \tau_{op}.
\label{Eq:EOMop}
\end{eqnarray}

There are two main driving situations. These correspond to driving around the frequencies of the two modes. We find it convenient to start with the case of driving around the optical mode frequency ($\omega_d\approx\omega_{op}$). We will come back to the case of driving around the acoustic mode frequency (with the drive frequency $\omega_d\approx\omega_{ac}$) later in the manuscript. In all of the cases that we consider in this work, to obtain significant contributions from both oscillators, we assume that the system is designed and/or biased such that $\omega_{op} \approx 2\omega_{ac}$.

A comment is in order here regarding the role of complex numbers in Eqs.~(\ref{Eq:EOMac}, \ref{Eq:EOMop}). It is common in the study of sinusoidally oscillating dynamical systems to start with real variables but then replace them with complex variables to simplify the algebraic manipulation of the equations. In Eqs.~(\ref{Eq:EOMac}, \ref{Eq:EOMop}), the variables are assumed to be complex from the beginning, as evidenced by the complex conjugation of $b_{ac}$ in Eq.~(\ref{Eq:EOMac}), as well as the imaginary factors $i$ in various terms. The complex-number nature of the variables in these equations has a physical interpretation: each one of the two magnetization variables ($b_{ac}$ and $b_{op}$) can be thought of as a two-dimensional magnetization vector, which can therefore be represented by a complex number. We will use units in which $b_{ac}$ and $b_{op}$ are dimensionless. As a result, $g_3$, $\tau_{ac}$ and $\tau_{op}$ will have units of frequency.

It is possible in principle to apply a rotating magnetic field that is described by the function $\tau_{op}=\tilde{\tau}_{op} \exp\left\{\pm i ( \omega_d t + \theta ) \right\}$. Alternatively, a linearly polarized sinusoidal driving field can be described by $\tau_{op}=\tilde{\tau}_{op} \cos(\omega_d t + \theta )$. The cosine function can then be expressed as a sum of two oppositely rotating components, one of which rotates in the same direction as the polarization and plays an important role in the dynamics, while the counter-rotating component can, to a good approximation, be ignored. Once we keep only one rotating field component, the time-dependence of $b_{ac}$ and $b_{op}$ can often be expressed through simple exponential functions with imaginary exponents and a single oscillation frequency for each variable, as we will see shortly.

An important question in the context of practical applications is what parameters can be tuned experimentally. The two mode frequencies are typically tuned by applying a static magnetic field. A typical situation is that in which the acoustic mode frequency increases while the optical mode frequency decreases with increasing magnetic field strength. For our purposes, to enhance the effect of inter-mode coupling on the dynamical response of the system, the field should be chosen close to the point where the acoustic mode frequency is about half the optical mode frequency. The driving (radio-frequency) field's amplitude, phase and polarization are also tunable parameters that affect the dynamical response. As shown in Refs.~\cite{Sud2025,Sud2026}, the polarization plays a role on how effectively the driving field couples to the modes. The frequency can be chosen close to the acoustic or optical mode. We will show below that the driven system exhibits qualitatively different dynamical responses in these two cases. We will also show that the driving field amplitude plays an important role in determining the steady state of the system and, as a result, the various features in the measured spectra.

\section{Driving near the optical mode frequency}
\label{Sec:Optical}

Let us take $\tau_{op}=\tilde{\tau}_{op} e^{-i\omega_d t}$ and assume that $\tilde{\tau}_{op}$ is real and that $\omega_d$ is near $\omega_{op}$. Since an electromagnetic drive signal often couples to both modes, we should in principle also take $\tau_{ac}=\tilde{\tau}_{ac} e^{-i\omega_d t}$. However, since this drive field is far off resonance with the acoustic mode frequency, we ignore it, i.e.~we set $\tau_{ac}=0$. If we now assume that $b_{ac}$ and $b_{op}$ oscillate in response to the drive field (all following exponential functions with imaginary exponents and single frequencies), and we require that all the terms in each one of the two equations [Eqs.~(\ref{Eq:EOMac},\ref{Eq:EOMop})] oscillate at the same frequency, we find that
\begin{eqnarray}
b_{ac} & = & c_{ac} e^{-i\tilde{\omega}_d t}, \\
b_{op} & = & \frac{c_{op}}{2} e^{-2i\tilde{\omega}_d t},
\end{eqnarray}
where $c_{ac}$ and $c_{op}$ are constants, and $\tilde{\omega}_d=\omega_d/2$. We have introduced the factors of 2 in the definitions of $c_{op}$ and $\tilde{\omega}_d$ for convenience, to make the next equations look more symmetric, even though this choice is not necessary in principle. Similarly, for convenience, we define the new parameters $\tilde{\omega}_{op}=\omega_{op}/2$ and $\tilde{\kappa}_{op}=\kappa_{op}/2$.

We can now easily use Eqs.~(\ref{Eq:EOMac}, \ref{Eq:EOMop}) to derive the equations that $c_{ac}$ and $c_{op}$ must satisfy:
\begin{equation}
\left(
\begin{array}{cc}
H_{11} & H_{21}^* \\
H_{21} & H_{22}
\end{array}
\right)
\left(
\begin{array}{cc}
c_{ac} \\
c_{op}
\end{array}
\right) =
\left(
\begin{array}{cc}
0 \\
\tilde{\tau}_{op}
\end{array}
\right),
\label{Eq:SteadyStateEquation}
\end{equation}
where
\begin{eqnarray}
H_{11} & = & \tilde{\omega}_d - \omega_{ac} + i \kappa_{ac}, \\
H_{22} & = & \tilde{\omega}_d - \tilde{\omega}_{op} + i \tilde{\kappa}_{op}, \\
H_{21} & = & i g_3 c_{ac}.
\end{eqnarray}
The above equations are very similar to Eq.~(3) in Ref.~\cite{Sud2025}, but with the driving field applied to the optical mode. We will see below that this difference leads to qualitative differences in the dynamics of the system. We will start by assuming that $g_3$ is real and positive. We will treat the case of imaginary $g_3$ later.

\subsection{Steady-state solutions}

We can now solve Eq.~(\ref{Eq:SteadyStateEquation}) to find the steady-state solutions. First we note that, since $H_{21}$ contains the factor $c_{ac}$, one simple solution that always exists is defined by $c_{ac}=0$, which then gives
\begin{equation}
c_{op} = \frac{\tilde{\tau}_{op}}{H_{22}} = \frac{\tilde{\tau}_{op}}{\tilde{\omega}_d - \tilde{\omega}_{op} + i \tilde{\kappa}_{op}}.
\label{Eq:copWhencac0}
\end{equation}
In other words, the optical mode reaches a steady state determined by a balance between driving and dissipation, while the acoustic mode remains in the static lowest-energy state, as if there is no driving at all. In fact, if the acoustic mode is not driven ($\tau_{ac}=0$), and $b_{ac}=0$ at the initial time, Eq.~(\ref{Eq:EOMac}) ensures that $b_{ac}$ remains equal to zero at all times.

Since Eq.~(\ref{Eq:SteadyStateEquation}) is nonlinear, it can have other solutions. We therefore look for solutions that have $c_{ac}\neq 0$. If any such solution exists, the top row in Eq.~(\ref{Eq:SteadyStateEquation}) gives
\begin{equation}
c_{op} = - \frac{H_{11}c_{ac}}{H_{21}^*}.
\label{Eq:copAsFnOfcac}
\end{equation}
Substituting Eq.~(\ref{Eq:copAsFnOfcac}) in the bottom row of Eq.~(\ref{Eq:SteadyStateEquation}) then gives
\begin{equation}
H_{21} c_{ac} - \frac{H_{11} H_{22} c_{ac}}{H_{21}^*} = \tilde{\tau}_{op}.
\end{equation}
Since $H_{21}$ contains the variable $c_{ac}$, we write it explicitly and [after multiplying the equation by $-ic_{ac}^*/(g_3 c_{ac})$] obtain
\begin{equation}
\left| c_{ac} \right|^2 - \frac{H_{11} H_{22}}{g_3^2} + i \frac{\tilde{\tau}_{op}}{g_3} \frac{c_{ac}^*}{c_{ac}} = 0.
\label{Eq:cacEquation}
\end{equation}
We now write $c_{ac} = \left| c_{ac} \right| e^{i\varphi}$ and [from the imaginary part of Eq.~(\ref{Eq:cacEquation})] obtain
\begin{equation}
\cos 2\varphi = \frac{\left( \tilde{\omega}_d - \omega_{ac} \right) \tilde{\kappa}_{op} + \left( \tilde{\omega}_d - \tilde{\omega}_{op} \right) \kappa_{ac}}{\tilde{\tau}_{op} g_3}.
\label{Eq:phiFormula}
\end{equation}
We now note that the sine can be positive or negative for the same value of the cosine: $\sin 2\varphi = \pm \sqrt{1-\cos^2 2\varphi}$. We therefore take the real part of Eq.~(\ref{Eq:cacEquation}) and obtain two possible solutions for $\left| c_{ac} \right|$:
\begin{widetext}
\begin{equation}
\left| c_{ac} \right|^2 = \frac{\left( \tilde{\omega}_d - \omega_{ac} \right) \left( \tilde{\omega}_d - \tilde{\omega}_{op} \right) - \kappa_{ac} \tilde{\kappa}_{op}}{g_3^2} \mp \sqrt{\frac{\tilde{\tau}_{op}^2}{g_3^2} - \left(\frac{\left( \tilde{\omega}_d - \omega_{ac} \right) \tilde{\kappa}_{op} + \left( \tilde{\omega}_d - \tilde{\omega}_{op} \right) \kappa_{ac}}{g_3^2}\right)^2}.
\label{Eq:cacFormula}
\end{equation}
\end{widetext}
In the following subsections, we will discuss the conditions under which the two solutions described by Eq.~(\ref{Eq:cacFormula}) are physical, as well as the dynamics close to the corresponding steady states.

\subsection{Conditions for the existence of $c_{ac}\neq 0$ steady-state solutions}

The fact that $\left| c_{ac} \right|$ must be real means that the $c_{ac}\neq 0$ steady-state solutions can exist only if certain conditions are satisfied. At a basic level, one can say that the right-hand side of Eq.~(\ref{Eq:cacFormula}) must be positive. However, Eq.~(\ref{Eq:cacFormula}) is rather long, which obscures its validity conditions. We therefore focus on different aspects of the formula separately. One condition can be derived by either requiring that the term inside the square root in Eq.~(\ref{Eq:cacFormula}) must be positive or requiring that $\cos 2\varphi$ in Eq.~(\ref{Eq:phiFormula}) must be between $-1$ and 1. The condition derived in this way reads:
\begin{equation}
\tilde{\tau}_{op} > \left| \frac{\left( \tilde{\omega}_d - \omega_{ac} \right) \tilde{\kappa}_{op} + \left( \tilde{\omega}_d - \tilde{\omega}_{op} \right) \kappa_{ac}}{g_3}\right|.
\label{Eq:ConditionPhysicalSine}
\end{equation}
When expressed in this way, the condition indicates that $c_{ac}\neq 0$ solutions can exist only when the driving field is sufficiently strong.

Assuming that the argument of the square-root in Eq.~(\ref{Eq:cacFormula}) is positive, we now try to develop further understanding of the necessary conditions by considering some special cases. First, let us assume that the decay rates are larger than the detunings. This scenario can be realized by setting the bias and driving conditions to $\omega_d\approx\omega_{op}\approx 2\omega_{ac}$. In this case, the first term in Eq.~(\ref{Eq:cacFormula}) is negative. We must therefore compare the first and second terms in Eq.~(\ref{Eq:cacFormula}) to determine whether the whole expression can be positive. Only the plus sign in Eq.~(\ref{Eq:cacFormula}) can give a valid solution. If we assume that the detunings are much smaller than the decay rates and can therefore be ignored, we obtain the approximate formula
\begin{equation}
\left| c_{ac} \right|^2 = -\frac{\kappa_{ac} \tilde{\kappa}_{op}}{g_3^2} + \frac{\tilde{\tau}_{op}}{g_3}.
\label{Eq:cacFormulaStrongDecay}
\end{equation}
The condition for the existence of this solution can therefore be derived as
\begin{equation}
\tilde{\tau}_{op} > \frac{\kappa_{ac} \tilde{\kappa}_{op}}{g_3}.
\end{equation}
Similarly to Eq.~(\ref{Eq:ConditionPhysicalSine}), this condition says that the driving strength must exceed a minimum threshold, noting here that this threshold is different from the one in Eq.~(\ref{Eq:ConditionPhysicalSine}).

Now let us assume that the decay rates are smaller than the detunings. In this case, in principle both signs in Eq.~(\ref{Eq:cacFormula}) can give valid solutions. For example, if the decay rates $\kappa_{ac}$ and $\kappa_{op}$ are both very small and can be ignored, and if we further consider the special case where $\left| \omega_{op} - 2\omega_{ac} \right| \ll \left| \omega_{d} - \omega_{op} \right|$, Eq.~(\ref{Eq:cacFormula}) can be approximated as
\begin{equation}
\left| c_{ac} \right|^2 \approx \frac{\left( \tilde{\omega}_d - \omega_{ac} \right)^2}{g_3^2} \mp \frac{\tilde{\tau}_{op}}{g_3}.
\label{Eq:cacFormulaWeakDecay}
\end{equation}
The plus sign always gives a valid solution. An additional solution (corresponding to the minus sign) exists if
\begin{equation}
\tilde{\tau}_{op} < \frac{\left( \tilde{\omega}_d - \omega_{ac} \right)^2}{g_3}.
\end{equation}
Interestingly, this condition says that the driving strength must be below a certain maximum value for the solution to exist.

\begin{figure}[h]
\includegraphics[width=5cm]{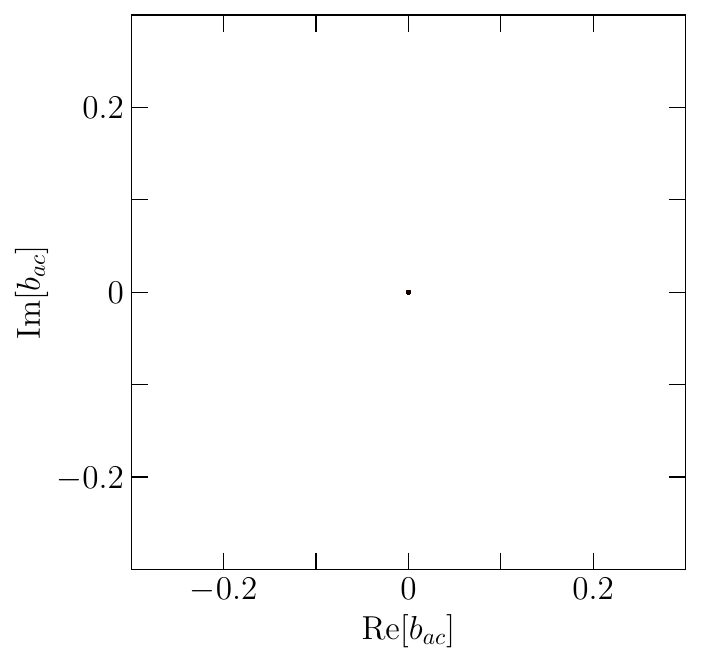}
\includegraphics[width=5cm]{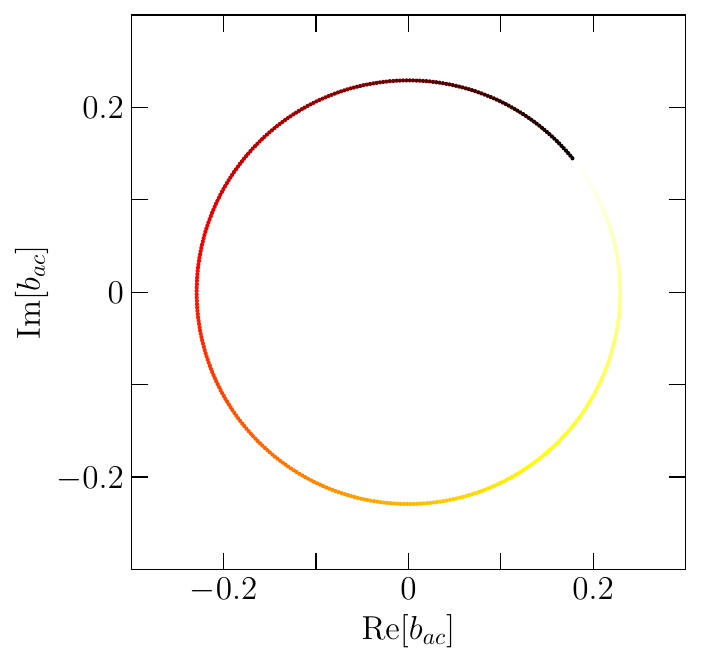}
\includegraphics[width=5cm]{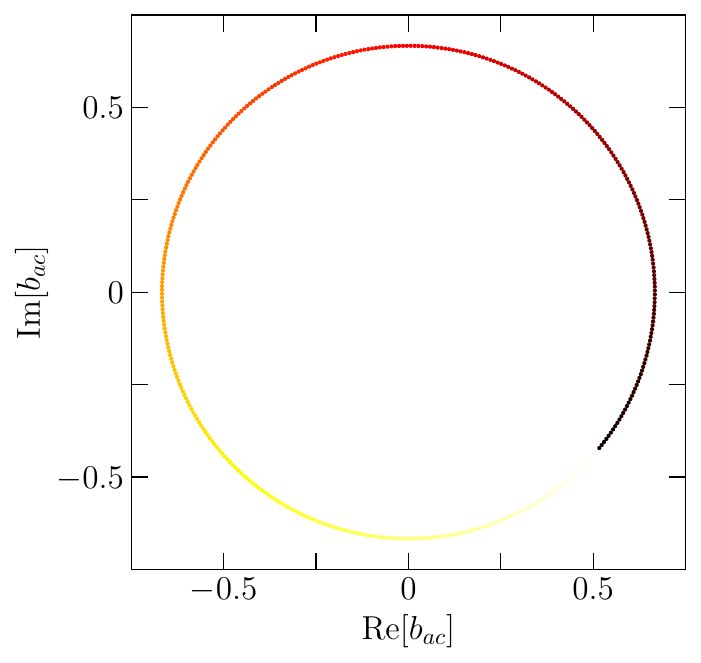}
\includegraphics[width=5cm]{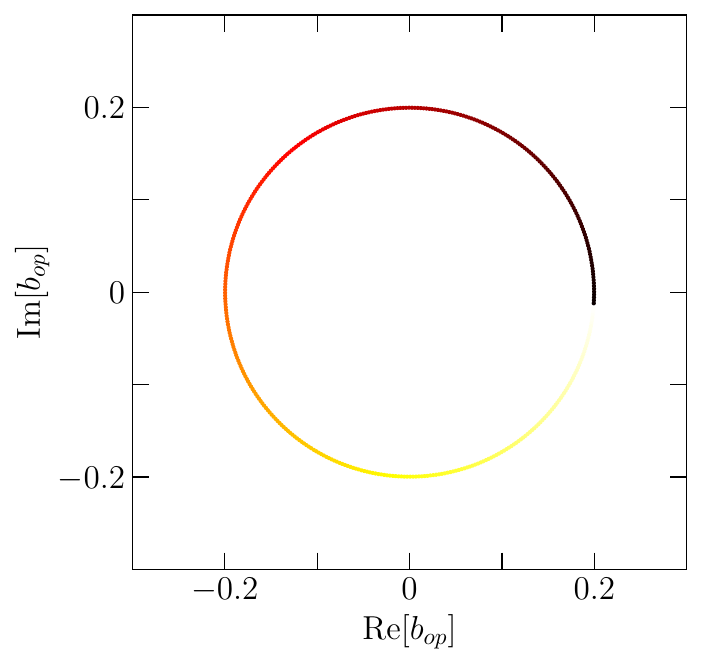}
\includegraphics[width=5cm]{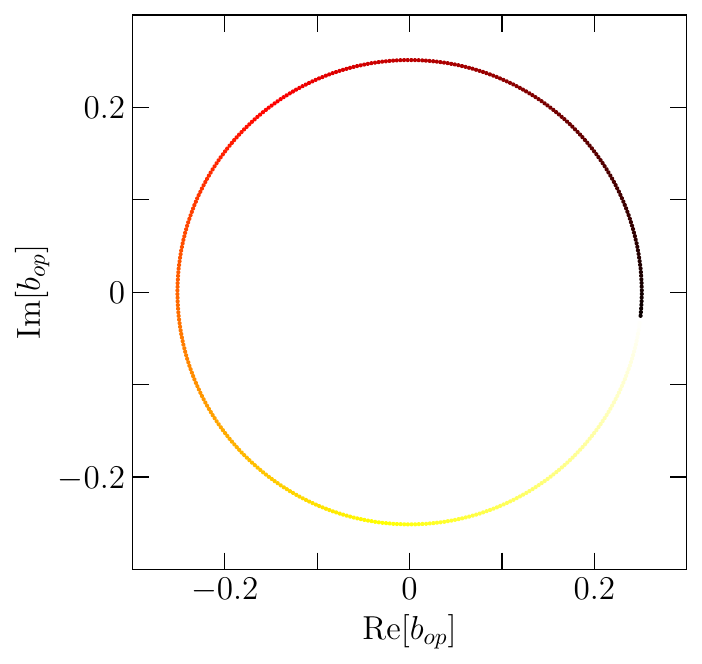}
\includegraphics[width=5cm]{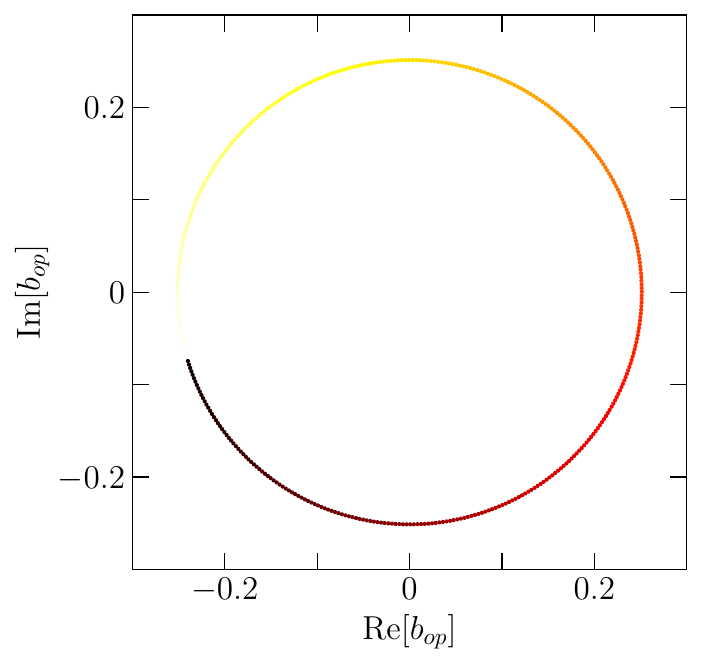}
\caption{Existence of three steady-state solutions. Each panel is a parametric plot showing the real and imaginary parts of one of the dynamical variables ($b_{ac}$ or $b_{op}$) as a function of time for one full oscillation period. In all cases, the system parameters are $\omega_{ac}=4$ GHz, $\omega_{op}=8$ GHz, $\omega_d=8.1$ GHz, $\kappa_{ac}=5$ MHz, $\kappa_{op}=6$ MHz, $g_3=0.1$ GHz, and $\tilde{\tau}_{op}=20$ MHz. The top row corresponds to $b_{ac}$, while the bottom row corresponds to $b_{op}$. The left column corresponds to the initial conditions $c_{ac}=0$ and $c_{op}$ given by Eq.~(\ref{Eq:copWhencac0}). The middle column corresponds to Eq.~(\ref{Eq:cacFormula}) with the minus sign. The right column corresponds to Eq.~(\ref{Eq:cacFormula}) with the plus sign. The color of each line evolves from white to black over one driving period. Numerical simulations show that among the three cases in this figure, only the one plotted in the right column is dynamically stable. In the other two cases, small perturbations grow and the system moves away from the the plotted curves (not shown in this figure).}
\label{Fig:ThreeSteadyStateSolutions}
\end{figure}

The general picture that emerges from the above conditions is as follows: There can be one, two or three steady-state solutions. Weak driving below a certain threshold gives only one solution. Strong driving above a certain threshold gives two solutions. This point can be seen by considering that the right-hand side of Eq.~(\ref{Eq:cacFormula}) can always be made positive with sufficiently large $\tilde{\tau}_{op}$. Intermediate strength driving can give three solutions, under suitable conditions. The existence of the three solutions is illustrated in Fig.~\ref{Fig:ThreeSteadyStateSolutions}.

\subsection{Normal modes near steady-state solutions}

Once we have identified a steady-state solution, we can investigate the dynamics of small deviations away from this solution. To do so, we express the dynamical variables $b_{ac}$ and $b_{op}$ as sums of the steady-state solution and additional small deviations from the steady-state solution:
\begin{eqnarray}
b_{ac} & = & c_{ac} e^{-i\tilde{\omega}_d t} + \delta_{ac}, \\
b_{op} & = & \frac{c_{op}}{2} e^{-i2\tilde{\omega}_d t} + \delta_{op},
\end{eqnarray}
where $\delta_{ac}$ and $\delta_{op}$ are the time-dependent small deviations. Substituting these equations into Eqs.~(\ref{Eq:EOMac},\ref{Eq:EOMop}) and eliminating the steady-state terms, we obtain the first-order approximate equations of motion for $\delta_{ac}$ and $\delta_{op}$:
\begin{eqnarray}
i \frac{d\delta_{ac}}{dt} & = & \left( \omega_{ac} - i \kappa_{ac} \right) \delta_{ac} + 2 i g_3 c_{ac}^* e^{i\tilde{\omega}_d t} \delta_{op} + i g_3 \delta_{ac}^* c_{op} e^{-2i\tilde{\omega}_d t},
\label{Eq:EOMPerturbationac}
\\
i \frac{d\delta_{op}}{dt} & = & \left( \omega_{op} - i \kappa_{op} \right) \delta_{op} - 2 i g_3 c_{ac} e^{-i\tilde{\omega}_d t} \delta_{ac}.
\label{Eq:EOMPerturbationop}
\end{eqnarray}
We now analyze the dynamics for the different steady-state solutions that we found in Sec.~\ref{Sec:Optical}.A.

\subsubsection{Case 1: $c_{ac}\neq 0$}

It is instructive to start with a seemingly natural approach that will turn out not to be the correct one. We look for oscillating-and-decaying solutions of the simplest possible form:
\begin{eqnarray}
\delta_{ac} & = & \delta_{ac,0} e^{-i \omega_1 t - \gamma_1 t},
\label{Eq:PerturbationAnsatz1ac}
\\
\delta_{op} & = & \delta_{op,0} e^{-i \omega_2 t - \gamma_2 t}.
\label{Eq:PerturbationAnsatz1op}
\end{eqnarray}
Substituting Eqs.~(\ref{Eq:PerturbationAnsatz1ac}) and (\ref{Eq:PerturbationAnsatz1op}) in Eqs.~(\ref{Eq:EOMPerturbationac}) and (\ref{Eq:EOMPerturbationop}), we obtain the equations
\begin{widetext}
\begin{eqnarray}
\left( \omega_1 - i \gamma_1 \right) \delta_{ac,0} e^{-i \omega_1 t - \gamma_1 t} & = & \left( \omega_{ac} - i \kappa_{ac} \right) \delta_{ac,0} e^{-i \omega_1 t - \gamma_1 t} + 2 i g_3 c_{ac}^* \delta_{op,0} e^{i(\tilde{\omega}_d - \omega_2) t - \gamma_2 t} \nonumber \\ & & + i g_3 \delta_{ac,0}^* c_{op} e^{i (-2\tilde{\omega}_d + \omega_1) t - \gamma_1 t},
%
\\
\left( \omega_2 - i \gamma_2 \right) \delta_{op,0} e^{-i \omega_2 t - \gamma_2 t} & = & \left( \omega_{op} - i \kappa_{op} \right) \delta_{op,0} e^{-i \omega_2 t - \gamma_2 t} - 2 i g_3 c_{ac} \delta_{ac,0} e^{-i (\tilde{\omega}_d + \omega_1) t - \gamma_1 t}.
%
\end{eqnarray}
\end{widetext}
If there are nontrivial (i.e.~nonzero) solutions to the above equations, the frequencies and decay rates of all the terms in each equation must be equal. If we require that this condition is satisfied, we find that we must have $\omega_1=\tilde{\omega}_d$, $\omega_2=2\tilde{\omega}_d$ and $\gamma_1=\gamma_2$ (which we denote as $\gamma$). If we make this substitution, we obtain the equations
\begin{widetext}
\begin{eqnarray}
\left( \tilde{\omega}_d - i \gamma \right) \delta_{ac,0} & = & \left( \omega_{ac} - i \kappa_{ac} \right) \delta_{ac,0} + 2 i g_3 c_{ac}^* \delta_{op,0} + i g_3 \delta_{ac,0}^* c_{op},
%
\\
\left( 2 \tilde{\omega}_d - i \gamma \right) \delta_{op,0} & = & \left( \omega_{op} - i \kappa_{op} \right) \delta_{op,0} - 2 i g_3 c_{ac} \delta_{ac,0},
%
\end{eqnarray}
\end{widetext}
which can be rewritten as four real equations:
\begin{equation}
\left(
\begin{array}{cccc}
h_{11} & h_{12} & h_{13} & h_{14} \\
h_{21} & h_{22} & h_{23} & h_{24} \\
h_{31} & h_{32} & h_{33} & h_{34} \\
h_{41} & h_{42} & h_{43} & h_{44}
\end{array}
\right)
\left(
\begin{array}{c}
{\rm Re} [ \delta_{ac,0} ] \\
{\rm Im} [ \delta_{ac,0} ] \\
{\rm Re} [ \delta_{op,0} ] \\
{\rm Im} [ \delta_{op,0} ]
\end{array}
\right) = 0,
\end{equation}
where
\begin{eqnarray}
h_{11} & = & \tilde{\omega}_d - \omega_{ac} + g_3 {\rm Im} [ c_{op} ], \\
h_{12} & = & \gamma - \kappa_{ac} - g_3 {\rm Re} [ c_{op} ], \\
h_{13} & = & - 2 g_3 {\rm Im} [ c_{ac} ], \\
h_{14} & = & 2 g_3 {\rm Re} [ c_{ac} ], \\
h_{21} & = & - \gamma + \kappa_{ac} - g_3 {\rm Re} [ c_{op} ], \\
h_{22} & = & \tilde{\omega}_d - \omega_{ac} - g_3 {\rm Im} [ c_{op} ], \\
h_{23} & = & - 2 g_3 {\rm Re} [ c_{ac} ], \\
h_{24} & = & - 2 g_3 {\rm Im} [ c_{ac} ], \\
h_{31} & = & - 2 g_3 {\rm Im} [ c_{ac} ], \\
h_{32} & = & - 2 g_3 {\rm Re} [ c_{ac} ], \\
h_{33} & = & 2 \tilde{\omega}_d - \omega_{op}, \\
h_{34} & = & \gamma - \kappa_{op}, \\
h_{41} & = & 2 g_3 {\rm Re} [ c_{ac} ], \\
h_{42} & = & - 2 g_3 {\rm Im} [ c_{ac} ], \\
h_{43} & = & - \gamma + \kappa_{op}, \\
h_{44} & = & 2 \tilde{\omega}_d - \omega_{op}.
\end{eqnarray}
For this equation to have nontrivial solutions, the determinant of the matrix must be zero. There is one unknown variable in the matrix, namely $\gamma$. It is therefore possible in principle that the determinant is zero for some values of $\gamma$, somewhat similarly to what happens in eigenvalue problems. We have performed numerical simulations to test the existence of solutions for this problem with a variety of parameter combinations. The results of our simulations suggest that no such solutions exist. Setting the determinant to zero and solving for $\gamma$, we consistently obtained four solutions, all of which contained both real and imaginary parts. The presence of the imaginary part indicates that the solutions are unphysical, since the problem has been transformed into a purely real one. Interestingly, the four obtained $\gamma$ values take the form of two pairs, each of which are complex conjugates of each other. The two distinct real parts of the different $\gamma$ values obtained in this way correspond to the two decay rates that we find in the solutions described below. The two distinct imaginary parts (ignoring the plus/minus sign differences) are almost equal, but we cannot identify their value with any frequency obtained in our numerical simulations of the dynamics.

We now turn to the approach that will lead to correct results. Instead of the single-term ans\"atze in Eqs.~(\ref{Eq:PerturbationAnsatz1ac}) and (\ref{Eq:PerturbationAnsatz1op}), we try multi-term ans\"atze of the form
\begin{eqnarray}
\delta_{ac} & = & \sum_{j=1}^{m} \delta_{ac,0,j} e^{-i \omega_{1,j} t - \gamma_{1,j} t},
\label{Eq:PerturbationAnsatz2ac}
\\
\delta_{op} & = & \sum_{j=1}^{m} \delta_{op,0,j} e^{-i \omega_{2,j} t - \gamma_{2,j} t}.
\label{Eq:PerturbationAnsatz2op}
\end{eqnarray}
Frequency matching analysis similar to that performed above for the single-term ans\"atze shows that two terms ($m=2$) are sufficient to satisfy the frequency-matching condition. Furthermore, if we take a solution with more terms ($m>2$), the solution can be broken down into smaller sets of terms that are decoupled from each other, as they will have different frequencies and/or decay rates. In this case, the condition that all the terms with the same frequency and decay rate must cancel each other gives $\omega_{1,1}+\omega_{1,2}=2\tilde{\omega}_d$ and $\omega_{2,j}-\omega_{1,j}=\tilde{\omega}_d$, with all the decay rates in Eqs.~(\ref{Eq:PerturbationAnsatz2ac}) and (\ref{Eq:PerturbationAnsatz2op}) equal to each other (with the single value $\gamma$).

With these relations taken into account, Eqs.~(\ref{Eq:EOMPerturbationac}) and (\ref{Eq:EOMPerturbationop}) give
\begin{widetext}
\begin{eqnarray}
\left( \omega_{1,j} - i \gamma \right) \delta_{ac,0,j} & = & \left( \omega_{ac} - i \kappa_{ac} \right) \delta_{ac,0,j} + 2 i g_3 c_{ac}^* \delta_{op,0,j} + i g_3 \delta_{ac,0,\overline{j}}^* c_{op},
\label{Eq:EigenEq2TPerturbationac}
\\
\left( \omega_{2,j} - i \gamma \right) \delta_{op,0,j} & = & \left( \omega_{op} - i \kappa_{op} \right) \delta_{op,0,j} - 2 i g_3 c_{ac} \delta_{ac,0,j},
\label{Eq:EigenEq2TPerturbationop}
\end{eqnarray}
\end{widetext}
where $\overline{j}=3-j$. Note that each equation above represents two separate equations (for $j=1,2$). Equation (\ref{Eq:EigenEq2TPerturbationop}) gives
\begin{equation}
\delta_{op,0,j} = \frac{- 2 i g_3 c_{ac} \delta_{ac,0,j}}{\left( \omega_{2,j} - \omega_{op} \right) - i \left(\gamma - \kappa_{op} \right)}.
\end{equation}
Substituting this formula into Eq.~(\ref{Eq:EigenEq2TPerturbationac}) gives
\begin{widetext}
\begin{equation}
\left[ \left( \omega_{1,j} - \omega_{ac} \right) - i \left(\gamma - \kappa_{ac} \right) - \frac{4 g_3^2 \left| c_{ac} \right|^2}{\left( \omega_{2,j} - \omega_{op} \right) - i \left(\gamma - \kappa_{op} \right)} \right] \delta_{ac,0,j} = i g_3 c_{op} \delta_{ac,0,\overline{j}}^*.
\end{equation}
Since $j$ has two values ($j=1,2$), the above equation can be expressed as
\begin{equation}
\left(
\begin{array}{cccc}
h_{11}^{(C)} & h_{12}^{(C)} \\
h_{21}^{(C)} & h_{22}^{(C)}
\end{array}
\right)
\left(
\begin{array}{c}
\delta_{ac,0,1} \\
\delta_{ac,0,2}^*
\end{array}
\right) = 0,
\label{Eq:EigenEq2TPerturbationComplex}
\end{equation}
where
\begin{eqnarray}
h_{11}^{(C)} & = & \left( \omega_{1,1} - \omega_{ac} \right) - i \left(\gamma - \kappa_{ac} \right) - \frac{4 g_3^2 \left| c_{ac} \right|^2}{\left( \omega_{2,1} - \omega_{op} \right) - i \left(\gamma - \kappa_{op} \right)}, \\
h_{12}^{(C)} & = & -i g_3 c_{op}, \\
h_{21}^{(C)} & = & i g_3 c_{op}^*, \\
h_{22}^{(C)} & = & \left( \omega_{1,2} - \omega_{ac} \right) + i \left(\gamma - \kappa_{ac} \right) - \frac{4 g_3^2 \left| c_{ac} \right|^2}{\left( \omega_{2,2} - \omega_{op} \right) + i \left(\gamma - \kappa_{op} \right)},
\end{eqnarray}
or alternatively
\begin{equation}
\left(
\begin{array}{cccc}
h_{11} & h_{12} & h_{13} & h_{14} \\
h_{21} & h_{22} & h_{23} & h_{24} \\
h_{31} & h_{32} & h_{33} & h_{34} \\
h_{41} & h_{42} & h_{43} & h_{44}
\end{array}
\right)
\left(
\begin{array}{c}
{\rm Re} [ \delta_{ac,0,1} ] \\
{\rm Im} [ \delta_{ac,0,1} ] \\
{\rm Re} [ \delta_{ac,0,2} ] \\
{\rm Im} [ \delta_{ac,0,2} ]
\end{array}
\right) = 0,
\label{Eq:EigenEq2TPerturbationReal}
\end{equation}
where
\begin{eqnarray}
h_{11} & = & \omega_{1,1} - \omega_{ac} - 4 g_3^2 \left| c_{ac} \right|^2 \frac{\left( \omega_{1,1} + \tilde{\omega}_d - \omega_{op} \right)}{\left( \omega_{1,1} + \tilde{\omega}_d - \omega_{op} \right)^2 + \left(\gamma - \kappa_{op} \right)^2}, \\
h_{12} & = & \gamma - \kappa_{ac} + 4 g_3^2 \left| c_{ac} \right|^2 \frac{\left(\gamma - \kappa_{op} \right)}{\left( \omega_{1,1} + \tilde{\omega}_d - \omega_{op} \right)^2 + \left(\gamma - \kappa_{op} \right)^2}, \\
h_{13} & = & g_3 {\rm Im} [ c_{op} ], \\
h_{14} & = & - g_3 {\rm Re} [ c_{op} ], \\
h_{21} & = & - \gamma + \kappa_{ac} - 4 g_3^2 \left| c_{ac} \right|^2 \frac{\left(\gamma - \kappa_{op} \right)}{\left( \omega_{1,1} + \tilde{\omega}_d - \omega_{op} \right)^2 + \left(\gamma - \kappa_{op} \right)^2}, \\
h_{22} & = & \omega_{1,1} - \omega_{ac} - 4 g_3^2 \left| c_{ac} \right|^2 \frac{\left( \omega_{1,1} + \tilde{\omega}_d - \omega_{op} \right)}{\left( \omega_{1,1} + \tilde{\omega}_d - \omega_{op} \right)^2 + \left(\gamma - \kappa_{op} \right)^2}, \\
h_{23} & = & - g_3 {\rm Re} [ c_{op} ], \\
h_{24} & = & - g_3 {\rm Im} [ c_{op} ], \\
h_{31} & = & g_3 {\rm Im} [ c_{op} ], \\
h_{32} & = &  - g_3 {\rm Re} [ c_{op} ], \\
h_{33} & = & 2 \tilde{\omega}_d - \omega_{1,1} - \omega_{ac} - 4 g_3^2 \left| c_{ac} \right|^2 \frac{\left( 3 \tilde{\omega}_d - \omega_{1,1} - \omega_{op} \right)}{\left( 3 \tilde{\omega}_d - \omega_{1,1} - \omega_{op} \right)^2 + \left(\gamma - \kappa_{op} \right)^2}, \\
h_{34} & = & \gamma - \kappa_{ac} + 4 g_3^2 \left| c_{ac} \right|^2 \frac{\left(\gamma - \kappa_{op} \right)}{\left( 3 \tilde{\omega}_d - \omega_{1,1} - \omega_{op} \right)^2 + \left(\gamma - \kappa_{op} \right)^2}, \\
h_{41} & = & g_3 {\rm Re} [ c_{op} ], \\
h_{42} & = & g_3 {\rm Im} [ c_{op} ], \\
h_{43} & = & \gamma - \kappa_{ac} + 4 g_3^2 \left| c_{ac} \right|^2 \frac{\left(\gamma - \kappa_{op} \right)}{\left( 3 \tilde{\omega}_d - \omega_{1,1} - \omega_{op} \right)^2 + \left(\gamma - \kappa_{op} \right)^2}, \\
h_{44} & = & \omega_{1,1} - 2 \tilde{\omega}_d + \omega_{ac} + 4 g_3^2 \left| c_{ac} \right|^2 \frac{\left( 3 \tilde{\omega}_d - \omega_{1,1} - \omega_{op} \right)}{\left( 3 \tilde{\omega}_d - \omega_{1,1} - \omega_{op} \right)^2 + \left(\gamma - \kappa_{op} \right)^2}.
\end{eqnarray}
\end{widetext}
Nontrivial solutions to Eq.~(\ref{Eq:EigenEq2TPerturbationReal}) [or Eq.~(\ref{Eq:EigenEq2TPerturbationComplex})] can exist only if the determinant of the $4 \times 4$ matrix vanishes. Considering that any analytical treatment of the problem is expected to lead to complicated algebra, we proceed with numerical treatment of specific examples.

\begin{figure}[h]
\includegraphics[width=5cm]{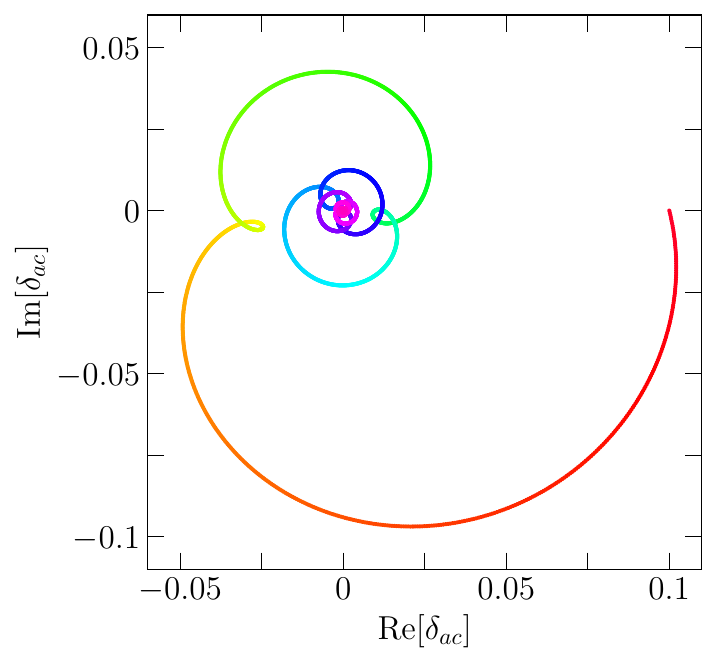}
\includegraphics[width=5cm]{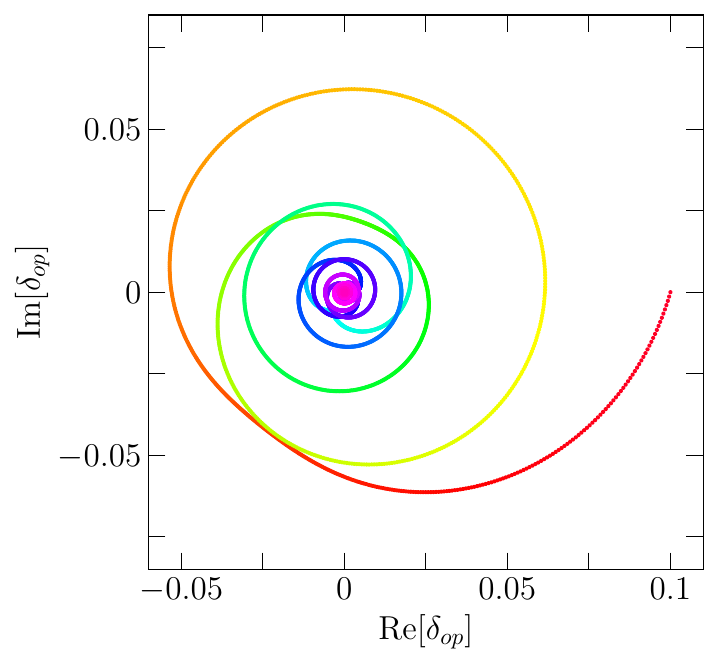}
\includegraphics[width=6cm]{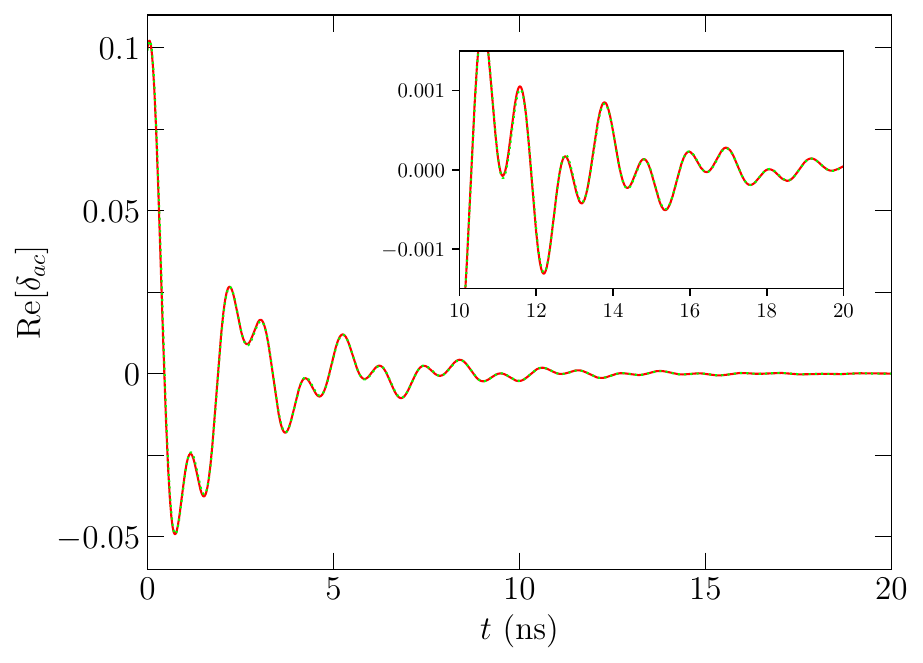}
\caption{Dynamics of small deviation from the steady state. Panels (a) and (b) are parametric plots each of which shows the real and imaginary parts of one of the dynamical variable deviations from the steady state (i.e.~$\delta_{ac}$ and $\delta_{op}$) as functions of time. The varying color is used to help the reader follow the trajectories of the lines. Panel (c) shows the real part of $\delta_{ac}$ as a function of time. The red solid line is obtained by solving the equations of motion, while the green dotted line is a theoretical fit. The frequencies and decay rates of the fitting function are calculated numerically from Eq.~(\ref{Eq:EigenEq2TPerturbationReal}), while the amplitudes and phases are obtained as fitting parameters optimized to fit the red line. The system parameters are $\omega_{ac}=4$ GHz, $\omega_{op}=8$ GHz, $\omega_d=8.1$ GHz, $\kappa_{ac}=0.5$ GHz, $\kappa_{op}=0.6$ GHz, $g_3=0.1$ GHz, and $\tilde{\tau}_{op}=10$ GHz. The initial conditions are $\delta_{ac}=\delta_{op}=0.1$.}
\label{Fig:PerturbationDynamics}
\end{figure}

First, we set the system parameters to $\omega_{ac}=4$ GHz, $\omega_{op}=8$ GHz, $\omega_d=8.1$ GHz, $\kappa_{ac}=0.5$ GHz, $\kappa_{op}=0.6$ GHz, $g_3=0.1$ GHz, and $\tilde{\tau}_{op}=10$ GHz. We calculate the steady-state solution for these parameters. The stable steady state with the chosen parameter set is described by Eq.~(\ref{Eq:cacFormula}) with the plus sign. We then set the initial conditions to $b_{ac}=c_{ac}+\delta_{ac}$ and $b_{op}=c_{op}/2+\delta_{op}$ with $\delta_{ac}=\delta_{op}=0.1$. We solve the equations of motion [Eqs.~(\ref{Eq:EOMac}) and (\ref{Eq:EOMop})] and subtract the steady-state solution to isolate the dynamics of the small deviations. The resulting dynamics are plotted in Fig.~\ref{Fig:PerturbationDynamics}. The plots show clearly that the dynamics contains multiple frequencies. We now calculate the normal mode frequencies from Eq.~(\ref{Eq:EigenEq2TPerturbationReal}). We obtain four pairs of $(\omega_{1,1},\gamma)$ values. These are (5.886, 0.79), (2.216, 0.79), (2.23, 0.311), and (5.872, 0.311). As expected, if we take the different $(\omega_{1,1},\gamma)$ pairs, we find that each one has a counterpart with the same decay rate ($\gamma$) and a complementary frequency such that $\omega_{1,1}+\omega_{1,2}=\omega_d$. As an example, we focus on the real part of $\delta_{ac}$ and construct a function with four terms, one term for each one of the $(\omega_{1,1},\gamma)$ pairs, along with a complex-number prefactor that serves as a fitting parameter in each term. In other words, the deviations are described by functions of the form
\begin{eqnarray}
\delta_{ac} & = & \sum_{j=1}^{4} \delta_{ac,0,j} e^{-i \omega_{1,j} t - \gamma_j t},
%
\\
\delta_{op} & = & \sum_{j=1}^{4} \delta_{op,0,j} e^{-i \omega_{2,j} t - \gamma_j t}.
\end{eqnarray}
The resulting functions provide excellent fit for the time dependence of $\delta_{ac}$, as shown in Fig.~\ref{Fig:PerturbationDynamics}. Similar results are obtained for all the other dynamical variables, i.e.~the real and imaginary parts of $\delta_{ac}$ and $\delta_{op}$.

\subsubsection{Case 2: $c_{ac}=0$}

As discussed in Sec.~\ref{Sec:Optical}, for the steady-state solution with $c_{ac}=0$, it follows that $c_{op}$ is given by Eq.~(\ref{Eq:copWhencac0}). In this case, Eqs.~(\ref{Eq:EOMPerturbationac}) and (\ref{Eq:EOMPerturbationop}) simplify to:
\begin{eqnarray}
i \frac{d\delta_{ac}}{dt} & = & \left( \omega_{ac} - i \kappa_{ac} \right) \delta_{ac} + i g_3 \delta_{ac}^* c_{op} e^{-2i\tilde{\omega}_d t},
\label{Eq:EOMPerturbationac0}
\\
i \frac{d\delta_{op}}{dt} & = & \left( \omega_{op} - i \kappa_{op} \right) \delta_{op}.
\label{Eq:EOMPerturbationop0}
\end{eqnarray}
Note that these two equations are decoupled: $\delta_{op}$ does not appear in Eq.~(\ref{Eq:EOMPerturbationac0}), and $\delta_{ac}$ does not appear in Eq.~(\ref{Eq:EOMPerturbationop0}).

Equation (\ref{Eq:EOMPerturbationop0}) immediately gives two modes with the same frequency ($\omega_{op}$) and the same decay rate ($\kappa_{op}$). The two modes can be taken as having real and imaginary initial amplitudes, such that any initial-state value of $\delta_{op}$ can be expressed as a sum of two components belonging to the two modes.

Following the same arguments as in the previous case (with $c_{ac}\neq 0$), we find that the solutions of Eq.~(\ref{Eq:EOMPerturbationac0}) can be expressed as:
\begin{equation}
\delta_{ac} = \sum_{j=1}^{2} \delta_{ac,0,j} e^{-i \omega_{1,j} t - \gamma t}.
\end{equation}
A simple frequency matching procedure gives $\omega_{1,1}+\omega_{1,2}=2\tilde{\omega}_d$. If we assume that $\omega_{1,1}\neq\omega_{1,2}$, we obtain the system of equations
\begin{widetext}
\begin{eqnarray}
\left( \omega_{1,1} - i \gamma \right) \delta_{ac,0,1} & = & \left( \omega_{ac} - i \kappa_{ac} \right) \delta_{ac,0,1} + i g_3 \delta_{ac,0,2}^* c_{op},
\\
\left( \omega_{1,2} - i \gamma \right) \delta_{ac,0,2} & = & \left( \omega_{ac} - i \kappa_{ac} \right) \delta_{ac,0,2} + i g_3 \delta_{ac,0,1}^* c_{op}.
\end{eqnarray}
The second equation gives
\begin{equation}
\delta_{ac,0,2} = \frac{i g_3 \delta_{ac,0,1}^* c_{op}}{\left( 2 \tilde{\omega}_d - \omega_{1,1} - \omega_{ac} \right) - i \left(\gamma - \kappa_{ac} \right)},
\end{equation}
which, upon substitution in the first equation, gives
\begin{equation}
\left[ \left( \omega_{1,1} - \omega_{ac} \right) - i \left(\gamma - \kappa_{ac} \right) - \frac{g_3^2 \left| c_{op} \right|^2}{\left( 2 \tilde{\omega}_d - \omega_{1,1} - \omega_{ac} \right) + i \left(\gamma - \kappa_{ac} \right)} \right] \delta_{ac,0,1} = 0.
\end{equation}
Nontrivial solutions (with $\delta_{ac,0,1}\neq 0$) can exist when
\begin{equation}
\left[ \left( \omega_{1,1} - \omega_{ac} \right) - i \left(\gamma - \kappa_{ac} \right) \right] \left[ \left( 2 \tilde{\omega}_d - \omega_{1,1} - \omega_{ac} \right) + i \left(\gamma - \kappa_{ac} \right) \right] - g_3^2 \left| c_{op} \right|^2 = 0.
\end{equation}
The imaginary part of this equation is
\begin{equation}
\left[ \left( \omega_{1,1} - \omega_{ac} \right) - \left( 2 \tilde{\omega}_d - \omega_{1,1} - \omega_{ac} \right) \right] \left(\gamma - \kappa_{ac} \right) = 0,
\end{equation}
which is satisfied when either $\omega_{1,1}=\tilde{\omega}_d$ (which also gives $\omega_{1,2}=\tilde{\omega}_d$) or $\gamma=\kappa_{ac}$. The real part of the equation is
\begin{equation}
\left( \omega_{1,1} - \omega_{ac} \right) \left( 2 \tilde{\omega}_d - \omega_{1,1} - \omega_{ac} \right) + \left(\gamma - \kappa_{ac} \right)^2 - g_3^2 \left| c_{op} \right|^2 = 0.
\label{Eq:ZeroCacSpectrumEquationRealPart}
\end{equation}
If $\gamma=\kappa_{ac}$, Eq.~(\ref{Eq:ZeroCacSpectrumEquationRealPart}) gives
\begin{equation}
\omega_{1,1} = \tilde{\omega}_d \pm \sqrt{\left(\tilde{\omega}_d - \omega_{ac} \right)^2 - g_3^2 \left| c_{op} \right|^2}.
\label{Eq:NormalModeFrequencyZeroCac}
\end{equation}
\end{widetext}

If $\omega_{1,1}=\omega_{1,2}=\tilde{\omega}_d$, we must first take a few steps back: in this case, the solutions of Eq.~(\ref{Eq:EOMPerturbationac0}) can be expressed as the single-term solution
\begin{equation}
\delta_{ac} = \delta_{ac,0} e^{-i \tilde{\omega}_d t - \gamma t}.
\end{equation}
Equation (\ref{Eq:EOMPerturbationac0}) then gives the system of equations
\begin{equation}
\left(
\begin{array}{cc}
h_{11} & h_{12} \\
h_{21} & h_{22}
\end{array}
\right)
\left(
\begin{array}{cc}
{\rm Re} [ \delta_{ac,0,1} ] \\
{\rm Im} [ \delta_{ac,0,1} ]
\end{array}
\right) = 0,
\label{Eq:Perturbationac0SingleTermEquation}
\end{equation}
with
\begin{eqnarray}
h_{11} & = & \tilde{\omega}_d - \omega_{ac} + g_3 {\rm Im} [ c_{op} ], \\
h_{12} & = & \gamma - \kappa_{ac} - g_3 {\rm Re} [ c_{op} ], \\
h_{21} & = & - \gamma + \kappa_{ac} - g_3 {\rm Re} [ c_{op} ], \\
h_{22} & = & \tilde{\omega}_d - \omega_{ac} - g_3 {\rm Im} [ c_{op} ].
\end{eqnarray}
Equation (\ref{Eq:Perturbationac0SingleTermEquation}) has nontrivial solutions when the determinant of the matrix vanishes, i.e.~when
\begin{equation}
\left( \tilde{\omega}_{d} - \omega_{ac} \right)^2 + \left( \gamma - \kappa_{ac} \right)^2 - g_3^2 \left| c_{op} \right|^2 = 0,
\end{equation}
or, in other words,
\begin{equation}
\gamma = \kappa_{ac} \pm \sqrt{g_3^2 \left| c_{op} \right|^2 - \left( \tilde{\omega}_{d} - \omega_{ac} \right)^2}.
\label{Eq:NormalModeDecayRateZeroCac}
\end{equation}
By inspecting the square roots in Eqs.~(\ref{Eq:NormalModeFrequencyZeroCac}) and (\ref{Eq:NormalModeDecayRateZeroCac}), we can see that one of them must give real values, while the other must give complex values. The physically relevant solution is the one that gives real values.

\subsection{Normal mode spectra}

\begin{figure}[h]
\includegraphics[width=5cm]{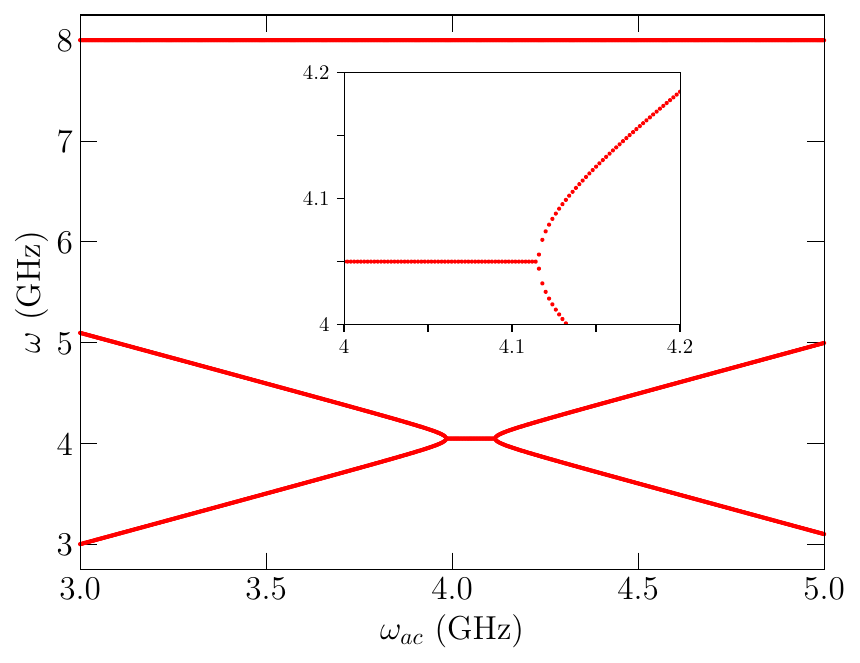}
\includegraphics[width=5cm]{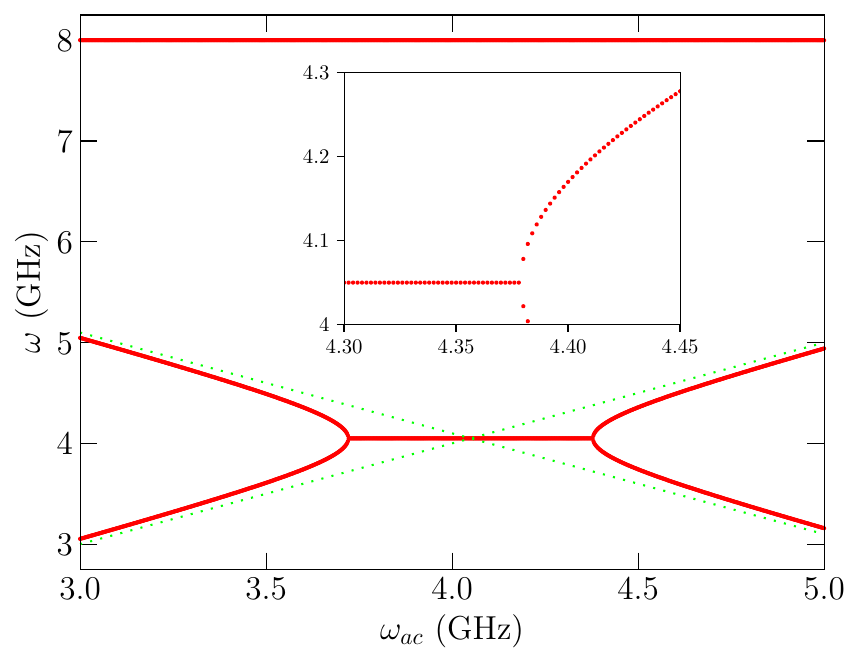}
\includegraphics[width=5cm]{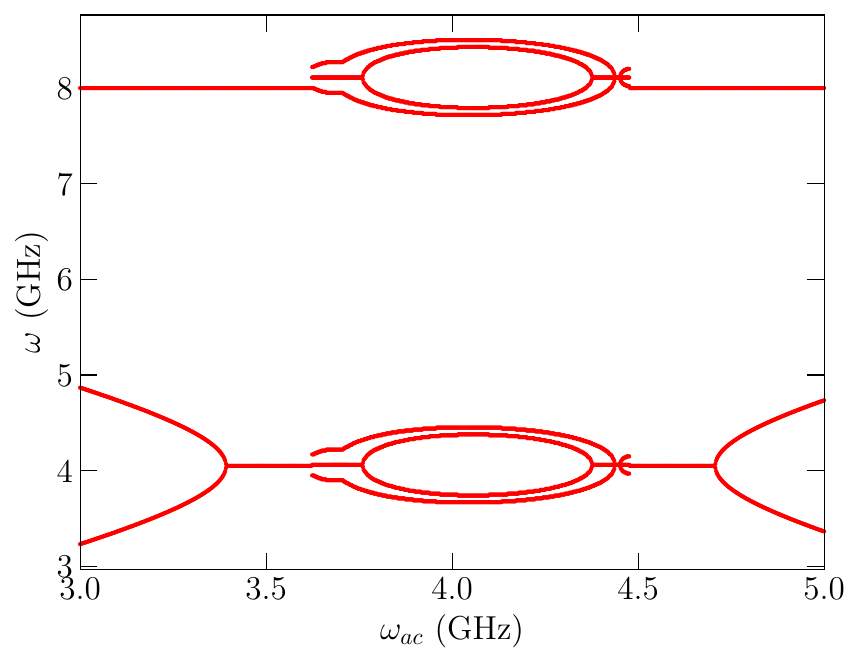}
\caption{Spectra of normal mode frequencies as functions of acoustic mode frequency ($\omega_{ac}$), with all other system parameters kept fixed. In all cases, we set $\omega_{op}=8$ GHz, $\omega_d=8.1$ GHz, $\kappa_{ac}=0.5$ GHz, $\kappa_{op}=0.6$ GHz, and $g_3=0.2$ GHz. In Panels (a), (b) and (c), $\tilde{\tau}_{op}=0.1, 0.5$ and 1 GHz, respectively. The insets zoom in on some of the exceptional points. The green dotted lines in Panel (b) are given by $\omega_{ac}$ and $\omega_d-\omega_{ac}$. They show the asymptotic behabiour of the red solid lines. For weak driving, only the $c_{ac}=0$ solution exists, and we obtain relatively simple level-attraction patterns. For strong driving (c) and $\omega_{op} \approx 2\omega_{ac}$, the steady-state switches to $c_{ac}\neq 0$, and a rather exotic structure appears in the spectrum.}
\label{Fig:SpectraDrivenOp}
\end{figure}

The rules described in the previous subsections can be combined to produce spectra of normal mode oscillation frequencies as functions of system parameters. This exercise can be used to predict or fit experimental data obtained by varying one of the system parameters and probing the response of the system at different frequency. A few examples of such spectra are shown in Fig.~\ref{Fig:SpectraDrivenOp}.

\subsection{Imaginary coupling constant}

We now consider the case of an imaginary coupling strength, which we obtain by replacing $g_3$ by $ig_3$ in Eqs.~(\ref{Eq:EOMac}) and (\ref{Eq:EOMop}), such that $g_3$ is still a real number.

Equations (\ref{Eq:EOMac}) and (\ref{Eq:EOMop}) are replaced by
\begin{eqnarray}
i \frac{db_{ac}}{dt} & = & \left( \omega_{ac} - i \kappa_{ac} \right) b_{ac} - 2 g_3 b_{ac}^* b_{op} + \tau_{ac},
%
\\
i \frac{db_{op}}{dt} & = & \left( \omega_{op} - i \kappa_{op} \right) b_{op} + g_3 b_{ac}^2 + \tau_{op}.
%
\end{eqnarray}
It turns out that these equations describe an unstable system. This situation can be seen by considering the case where $b_{ac}$ is a large real number and $b_{op}$ is a large real number multiplied by $-i$. In this case, both variables grow rapidly towards infinity. Our numerical simulations indeed show unstable dynamics ($b_{ac}$ and $b_{op}$ diverging to $\infty$) when the driving fields $\tau_{ac/op}$ and/or the initial amplitudes $b_{ac/op}$ are sufficiently large. Nevertheless, the $c_{ac}=0$ solution is dynamically stable for some parameters. We therefore analyze this case here.

The equation for the steady-state solution now reads
\begin{equation}
\left(
\begin{array}{cc}
H_{11} & -H_{21}^* \\
H_{21} & H_{22}
\end{array}
\right)
\left(
\begin{array}{cc}
c_{ac} \\
c_{op}
\end{array}
\right) =
\left(
\begin{array}{cc}
0 \\
\tilde{\tau}_{op}
\end{array}
\right),
%
\end{equation}
with
\begin{eqnarray}
H_{11} & = & \tilde{\omega}_d - \omega_{ac} + i \kappa_{ac}, \\
H_{22} & = & \tilde{\omega}_d - \tilde{\omega}_{op} + i \tilde{\kappa}_{op}, \\
H_{21} & = & - g_3 c_{ac}.
\end{eqnarray}
Since the off-diagonal matrix elements do not affect the formula for the $c_{ac}=0$ solution, this solution is still given by
\begin{equation}
c_{op} = \frac{\tilde{\tau}_{op}}{H_{22}} = \frac{\tilde{\tau}_{op}}{\tilde{\omega}_d - \tilde{\omega}_{op} + i \tilde{\kappa}_{op}}.
%
\end{equation}

The dynamics of small deviations away from the steady state are now described by the equations
\begin{eqnarray}
i \frac{d\delta_{ac}}{dt} & = & \left( \omega_{ac} - i \kappa_{ac} \right) \delta_{ac} - g_3 \delta_{ac}^* c_{op} e^{-2i\tilde{\omega}_d t},
\label{Eq:EOMPerturbationac0ImG}
\\
i \frac{d\delta_{op}}{dt} & = & \left( \omega_{op} - i \kappa_{op} \right) \delta_{op}.
\label{Eq:EOMPerturbationop0ImG}
\end{eqnarray}
As in the case of real coupling strength, these two equations are decoupled: $\delta_{op}$ does not appear in Eq.~(\ref{Eq:EOMPerturbationac0ImG}), and $\delta_{ac}$ does not appear in Eq.~(\ref{Eq:EOMPerturbationop0ImG}). Similarly, Eq.~(\ref{Eq:EOMPerturbationop0ImG}) still gives two modes with the same frequency ($\omega_{op}$) and the same decay rate ($\kappa_{op}$).

The solutions of Eq.~(\ref{Eq:EOMPerturbationac0ImG}) can be expressed as:
\begin{equation}
\delta_{ac} = \sum_{j=1}^{2} \delta_{ac,0,j} e^{-i \omega_{1,j} t - \gamma t}.
\end{equation}
Frequency matching gives $\omega_{1,1}+\omega_{1,2}=2\tilde{\omega}_d$. If we assume that $\omega_{1,1}\neq\omega_{1,2}$, we obtain the system of equations
\begin{widetext}
\begin{eqnarray}
\left( \omega_{1,1} - i \gamma \right) \delta_{ac,0,1} & = & \left( \omega_{ac} - i \kappa_{ac} \right) \delta_{ac,0,1} - g_3 \delta_{ac,0,2}^* c_{op},
\\
\left( \omega_{1,2} - i \gamma \right) \delta_{ac,0,2} & = & \left( \omega_{ac} - i \kappa_{ac} \right) \delta_{ac,0,2} - g_3 \delta_{ac,0,1}^* c_{op}.
\end{eqnarray}
The second equation gives
\begin{equation}
\delta_{ac,0,2} = \frac{- g_3 \delta_{ac,0,1}^* c_{op}}{\left( 2 \tilde{\omega}_d - \omega_{1,1} - \omega_{ac} \right) - i \left(\gamma - \kappa_{ac} \right)},
\end{equation}
which, upon substitution in the first equation, gives
\begin{equation}
\left[ \left( \omega_{1,1} - \omega_{ac} \right) - i \left(\gamma - \kappa_{ac} \right) - \frac{g_3^2 \left| c_{op} \right|^2}{\left( 2 \tilde{\omega}_d - \omega_{1,1} - \omega_{ac} \right) + i \left(\gamma - \kappa_{ac} \right)} \right] \delta_{ac,0,1} = 0.
\end{equation}
\end{widetext}
This is the same equation as in the case of a real coupling constant. Therefore, the spectrum will behave in the same way as that discussed in Sec.~\ref{Sec:Optical}.D, as long as the point $c_{ac}=0$ is dynamically stable and the deviation from the steady state is small.

\section{Driving near the acoustic mode frequency}
\label{Sec:Acoustic}

When the driving field is close to resonance with the acoustic mode, we can make the approximation that the driving field couples only to this mode. The appropriate expressions for the driving term and steady-state dynamical variables in this case are
\begin{eqnarray}
\tau_{ac} & = & \tilde{\tau}_{ac} e^{-i\omega_d t}, \\
b_{ac} & = & c_{ac} e^{-i\omega_d t}, \\
b_{op} & = & \frac{c_{op}}{2} e^{-2i\omega_d t}.
\end{eqnarray}
With these expressions, the equations for the steady-state solutions read
\begin{equation}
\left(
\begin{array}{cc}
H_{11} & H_{21}^* \\
H_{21} & H_{22}
\end{array}
\right)
\left(
\begin{array}{cc}
c_{ac} \\
c_{op}
\end{array}
\right) =
\left(
\begin{array}{cc}
\tilde{\tau}_{ac} \\
0
\end{array}
\right),
\label{Eq:SteadyStateEquationACDrive}
\end{equation}
with the matrix elements $H_{ij}$ having the same form as before, but with $\tilde{\omega}_d$ replaced by $\omega_d$. In other words, the most significant change is in the right-hand side of the equation.

The bottom row of Eq.~(\ref{Eq:SteadyStateEquationACDrive}) gives
\begin{equation}
c_{op} = - \frac{H_{21}c_{ac}}{H_{22}},
%
\end{equation}
which, upon substitution in the top row of Eq.~(\ref{Eq:SteadyStateEquationACDrive}), gives
\begin{equation}
H_{11} c_{ac} - \frac{g_3^2 \left|c_{ac}\right|^2 c_{ac}}{H_{22}} = \tilde{\tau}_{ac}.
\label{Eq:cacEquationDriveAc}
\end{equation}
As before, this equation is nonlinear and can allow multiple solutions. However, unlike the case of driving the optical mode, there are generally no solution with $c_{ac}=0$ or $c_{op}=0$ when $\tilde{\tau}_{ac}\neq 0$. Furthermore, we cannot simplify the problem by evaluating the absolute value and phase of $c_{ac}$ separately using the real and imaginary parts of the equation. We can, however, obtain approximate expressions for $c_{ac}$ for weak driving.

If we set $\tilde{\tau}_{ac}=0$, we can immediately see that Eq.~(\ref{Eq:cacEquationDriveAc}) always has the solutions $c_{ac}=0$. The other solution, or rather infinite set of solutions defined by the equation
\begin{equation}
\left|c_{ac}\right|^2 = \frac{g_3^2}{H_{11} H_{22}},
\end{equation}
exist only when $H_{11} H_{22}$ is real. This condition can in principle be satisfied for one specific drive frequency that lies between $\omega_{ac}$ and $\tilde{\omega}_{op}$.

If we focus on the one solution that approaches $c_{ac}=0$ when $\tilde{\tau}_{ac}$ approaches zero, we can express it as
\begin{equation}
c_{ac} = \frac{\tilde{\tau}_{ac}}{\left( H_{11} - \frac{g_3^2 \left|c_{ac}\right|^2}{H_{22}} \right)}.
\label{Eq:cacEquationDriveAc}
\end{equation}
To lowest approximation, this equation gives $c_{ac} \approx \tilde{\tau}_{ac}/H_{11}$, and better approximations can be obtained by substituting an approximate expression for $c_{ac}$ in the denominator on the right-hand side of the equation.

\begin{figure}[h]
\includegraphics[width=5cm]{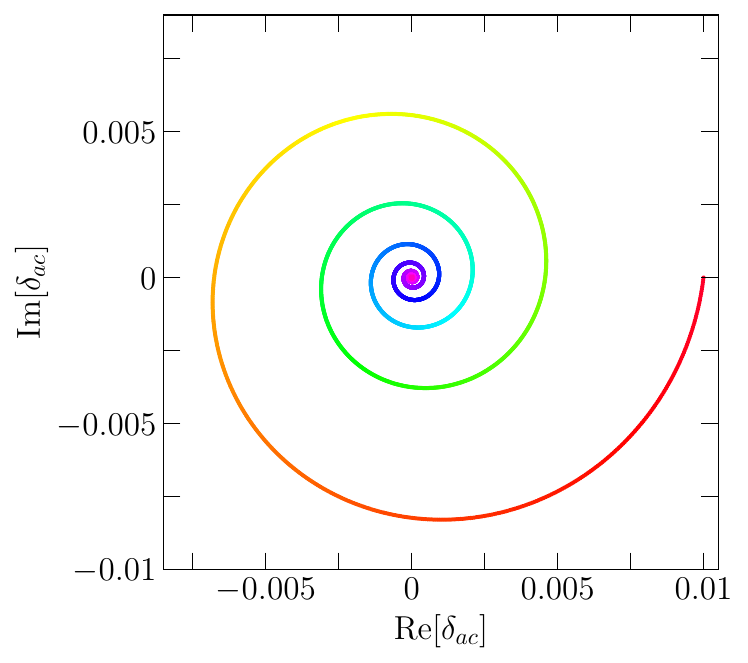}
\includegraphics[width=5cm]{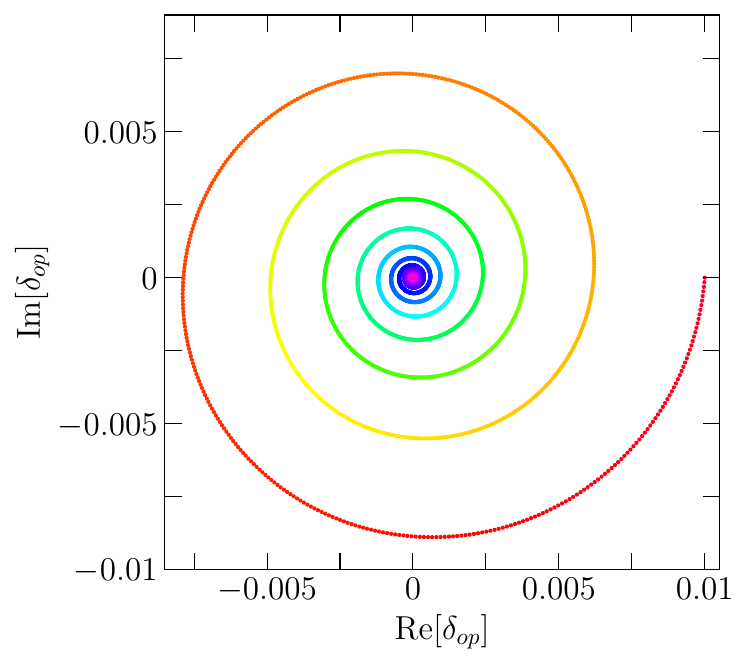}
\includegraphics[width=6cm]{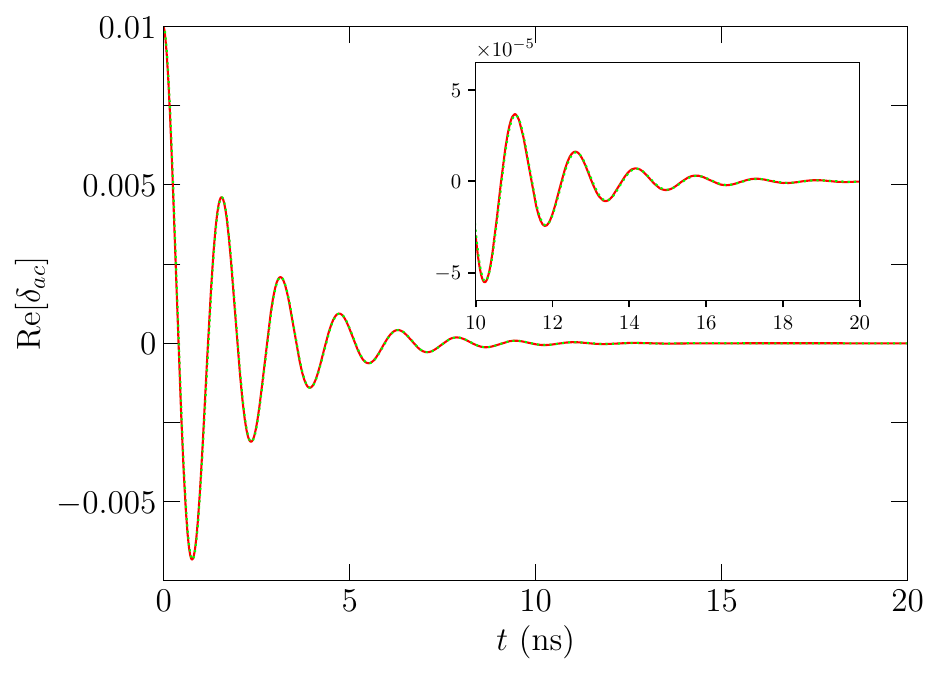}
\caption{Dynamics of small deviation from the steady state when the driving field is applied to the acoustic mode. The system parameters are $\omega_{ac}=4$ GHz, $\omega_{op}=8$ GHz, $\omega_d=4.1$ GHz, $\kappa_{ac}=0.5$ GHz, $\kappa_{op}=0.6$ GHz, $g_3=0.1$ GHz, and $\tilde{\tau}_{ac}=0.1$ GHz. The initial conditions are $\delta_{ac}=\delta_{op}=0.01$. Panels (a) and (b) are parametric plots each of which shows the real and imaginary parts of one of the dynamical variable deviations from the steady state (i.e.~$\delta_{ac}$ and $\delta_{op}$) as functions of time. The same variable color is used as in Fig.~\ref{Fig:PerturbationDynamics}. Panel (c) shows the real part of $\delta_{ac}$ as a function of time. As in Fig.~\ref{Fig:PerturbationDynamics}, the red solid line is obtained by solving the equations of motion, while the green dotted line is a theoretical fit. Unlike the case of driving the optical mode, the oscillations look mostly like simple damped oscillations. Nevertheless, to obtain a good fit we must include two oscillating-and-decaying terms in the fitting function, which is consistent with the derivations in the main text.}
\label{Fig:PerturbationDynamicsDriveAc}
\end{figure}

\begin{figure}[h]
\includegraphics[width=5cm]{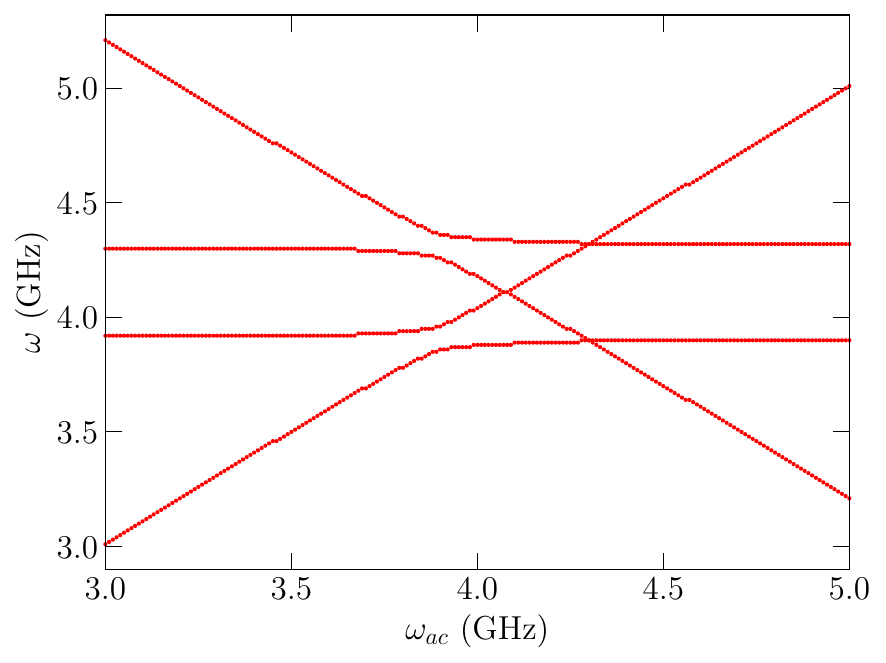}
\includegraphics[width=5cm]{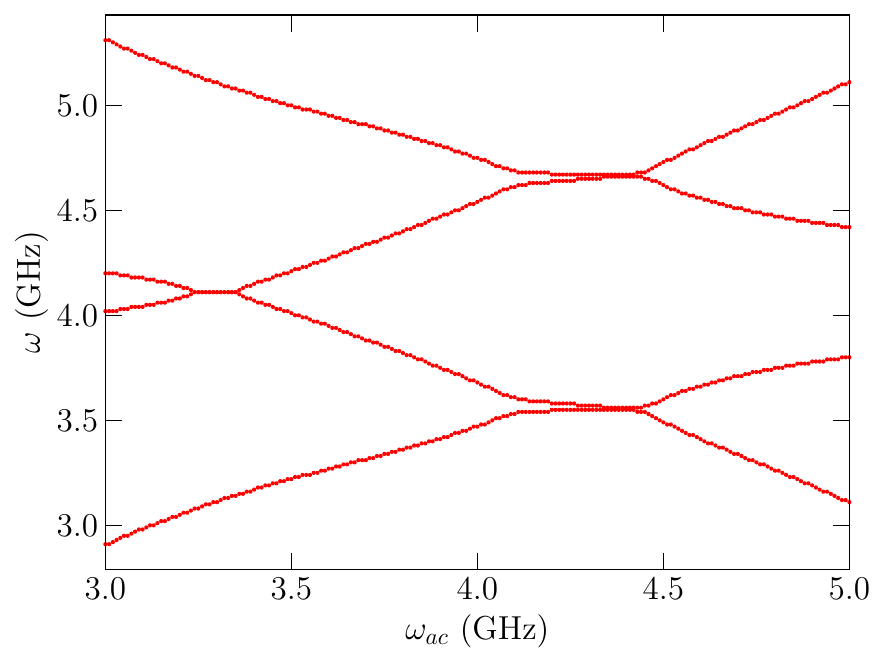}
\caption{Spectra of normal mode frequencies as functions of acoustic mode frequency ($\omega_{ac}$), with all other system parameters kept fixed, when the driving field is applied to the acoustic mode. In both panels, we set $\omega_{op}=8$ GHz, $\omega_d=4.1$ GHz, $\kappa_{ac}=0.5$ GHz, $\kappa_{op}=0.6$ GHz, and $g_3=0.2$ GHz. In Panels (a) and (b), $\tilde{\tau}_{ac}=0.1$ and 1 GHz, respectively. Note that for every frequency $\omega$ plotted here, there is another frequency at $\omega+\omega_d$ that is not shown in this figure. For weak driving (a), we obtain a simple level-repulsion pattern. For strong driving, we obtain a more complicated spectrum in which the level-repulsion pattern is obscured by other features that are likely due to the complicated dependence of $c_{ac}$ on $\tilde{\tau}_{ac}$. Furthermore, level-attraction patterns appear in the spectrum.}
\label{Fig:SpectraDrivenAc}
\end{figure}

The equations of motion for small deviations away from the steady state are given by
\begin{eqnarray}
i \frac{d\delta_{ac}}{dt} & = & \left( \omega_{ac} - i \kappa_{ac} \right) \delta_{ac} + 2 i g_3 c_{ac}^* e^{i\omega_d t} \delta_{op} + i g_3 \delta_{ac}^* c_{op} e^{-2i\omega_d t},
\label{Eq:EOMPerturbationacDriveAc}
\\
i \frac{d\delta_{op}}{dt} & = & \left( \omega_{op} - i \kappa_{op} \right) \delta_{op} - 2 i g_3 c_{ac} e^{-i\omega_d t} \delta_{ac}.
\label{Eq:EOMPerturbationopDriveAc}
\end{eqnarray}
The subsequent analysis of the oscillation modes then follows closely that presented in Sec.~\ref{Sec:Optical}.C, with the main difference being that the formulae for $c_{ac}$ and $c_{op}$ are different. In spite of this similarity, the response of the system to the driving field turns out to be significantly different. In particular, for weak driving, the spectra of normal mode frequencies now exhibit level-repulsion patterns, instead of the level-attraction patterns that we found in Sec.~\ref{Sec:Optical}. Examples of the dynamics and spectra are shown in Figs.~\ref{Fig:PerturbationDynamicsDriveAc} and \ref{Fig:SpectraDrivenAc}.

For weak driving, we obtain a simple level-repulsion pattern in the spectrum. This feature can be explained intuitively rather easily. If we make the observation that, for small $\tilde{\tau}_{ac}$, $c_{ac}$ is proportional to $\tilde{\tau}_{ac}$ and $c_{op}$ is proportional to $\tilde{\tau}_{ac}^2$, we can make the approximation that the last term in Eq.~(\ref{Eq:EOMPerturbationacDriveAc}) is negligible. This approximation was made implicitly in Ref.~\cite{Sud2025}. Under this approximation, it becomes rather obvious that we will obtain spectra with avoided-level patterns. Interestingly, for the case of strong driving shown in Fig.~\ref{Fig:SpectraDrivenAc}, we do not see any level-repulsion patterns. Instead, we obtain a few level-attraction patterns. The two such features around $\omega_{ac}=4.4$ GHz are rather unusual in that the lines seem to merge smoothly on the left but bifurcate more abruptly on the right.

It is worth pausing here to comment on the difference between the case of weakly driving the optical mode, where we obtained level attraction patterns in the spectrum, and the case of weakly driving the acoustic mode, where we obtained level repulsion patterns. The difference lies in the difference between Eq.~(\ref{Eq:EOMPerturbationac0}) and Eqs.~(\ref{Eq:EOMPerturbationacDriveAc},\ref{Eq:EOMPerturbationopDriveAc}), which govern the dynamics of small deviations from the steady state. The difference between these equations is a result of the asymmetric form of the nonlinear interaction terms in the original equations of motion, Eqs.~(\ref{Eq:EOMac},\ref{Eq:EOMop}). In the case of Eq.~(\ref{Eq:EOMPerturbationac0}), the linearized equation of motion for the small deviations from the steady state contain a coupling term between $\delta_{ac}$ and $\delta_{ac}^*$, which leads to level attraction. In contrast, when dealing with Eqs.~(\ref{Eq:EOMPerturbationacDriveAc},\ref{Eq:EOMPerturbationopDriveAc}), if we ignore the last term in Eq.~(\ref{Eq:EOMPerturbationacDriveAc}), we obtain Hermitian coupling between $\delta_{ac}$ and $\delta_{op}$, which leads to level repulsion.

\section{Conclusion}
\label{Sec:Conclusion}

We have analyzed the dynamical behaviour of two driven dissipative oscillators with nonlinear, non-Hermitian coupling. We identified the steady states under various system parameter combinations and driving conditions. We then analyzed the dynamics of small deviations from the steady states. This analysis allows us to obtain the normal mode spectra under the different conditions that we considered. We found a variety of spectral patterns under different circumstances, including level-repulsion and level-attraction patterns in the spectra. Our theoretical results explain recent experimental measurements on synthetic antiferromagnets, and they predict the existence of other types of spectra under different driving circumstances. These spectra, which have never been seen in experiment, could be observed in future experiments based on our theoretical predictions.

\section*{Acknowledgment}

This work was supported by Japan's Ministry of Education, Culture, Sports, Science and Technology (MEXT) Quantum Leap Flagship Program Grant Number JPMXS0120319794, and the Japan Society for the Promotion of Science (JSPS) Kakenhi Grant Numbers 24H02235 and 25K24642.

\appendix

\section{Two linearly coupled oscillators}

To gain additional insight into the roles of different elements in the non-linear problem under study in this work, we apply the procedure used in the main text to a simpler problem. We take two coupled linear equations, with a driving force applied to the first equation:
\begin{eqnarray}
i \frac{db_{1}}{dt} & = & \left( \omega_{1} - i \kappa_{1} \right) b_{1} + g b_{2} + \tau e^{-i\omega_d t},
\label{Eq:EOM_Lin1}
\\
i \frac{db_{2}}{dt} & = & \left( \omega_{2} - i \kappa_{2} \right) b_{2} + g b_{1}.
\label{Eq:EOM_Lin2}
\end{eqnarray}
In principle, our derivations below do not depend on whether $g$ is real or complex. We will see at the end of this section that real $g$ can be thought of as the case of coherent coupling, while imaginary $g$ can be thought of as the case of dissipative coupling.

The dynamical variables $b_{1}$ and $b_{2}$ oscillate in response to the driving field. Since all the terms other than the driving term are linear in either $b_{1}$ or $b_{2}$, the frequency matching condition implies that the steady state is described by the functions
\begin{eqnarray}
b_{1} & = & c_{1} e^{-i\omega_d t}, \\
b_{2} & = & c_{2} e^{-i\omega_d t},
\end{eqnarray}
where $c_{1}$ and $c_{2}$ are constants.

We can now easily derive the equations that $c_{1}$ and $c_{2}$ must satisfy:
\begin{equation}
\left(
\begin{array}{cc}
H_{11} & H_{21} \\
H_{21} & H_{22}
\end{array}
\right)
\left(
\begin{array}{cc}
c_{1} \\
c_{2}
\end{array}
\right) =
\left(
\begin{array}{cc}
\tau \\
0
\end{array}
\right),
\label{Eq:SteadyStateEquation_Lin}
\end{equation}
where
\begin{eqnarray}
H_{11} & = & \omega_d - \omega_{1} + i \kappa_{1}, \\
H_{22} & = & \omega_d - \omega_{2} + i \kappa_{2}, \\
H_{21} & = & g.
\end{eqnarray}

The bottom row of Eq.~(\ref{Eq:SteadyStateEquation_Lin}) gives
\begin{equation}
c_{2} = - \frac{H_{21}}{H_{22}} c_1,
%
\end{equation}
which, upon substitution in the top row of Eq.~(\ref{Eq:SteadyStateEquation_Lin}), gives
\begin{equation}
\left( H_{11} - \frac{H_{21}^2}{H_{22}} \right) c_{1} = \tau,
\end{equation}
which then gives
\begin{eqnarray}
c_{1} & = & \frac{\tau}{\left( \omega_d - \omega_{1} + i \kappa_{1} \right) - g^2 / \left( \omega_d - \omega_{2} + i \kappa_{2} \right)}, \\
c_{2} & = & - \frac{g\tau}{\left( \omega_d - \omega_{1} + i \kappa_{1} \right) \left( \omega_d - \omega_{2} + i \kappa_{2} \right) - g^2}.
\label{Eq:cFormula_Lin}
\end{eqnarray}

The normal modes are now found by adding small deviations from the steady-state solution:
\begin{eqnarray}
b_{1} & = & c_{1} e^{-i\omega_d t} + \delta_{1}, \\
b_{2} & = & c_{2} e^{-i\omega_d t} + \delta_{2},
\end{eqnarray}
where $\delta_{1}$ and $\delta_{2}$ are the time-dependent small deviations. Substituting these equations into Eqs.~(\ref{Eq:EOM_Lin1},\ref{Eq:EOM_Lin2}) and eliminating the steady-state terms, we obtain the first-order approximate equations of motion for $\delta_{1}$ and $\delta_{2}$:
\begin{eqnarray}
i \frac{d\delta_{1}}{dt} & = & \left( \omega_{1} - i \kappa_{1} \right) \delta_{1} + g \delta_{2},
\label{Eq:EOMPerturbation_Lin1}
\\
i \frac{d\delta_{2}}{dt} & = & \left( \omega_{2} - i \kappa_{2} \right) \delta_{2} + g \delta_{1}.
\label{Eq:EOMPerturbation_Lin2}
\end{eqnarray}
Note that the driving force ($\tau$) does not appear in these equations, which means that the normal modes do not depend on the details of the driving force. This situation is indeed what we expect in a linear system.

Since the equations are linear, we look for oscillating-and-decaying solutions of the simple form:
\begin{eqnarray}
\delta_{1} & = & \delta_{1,0} e^{-i \omega t - \gamma t},
\label{Eq:PerturbationAnsatz_Lin1}
\\
\delta_{2} & = & \delta_{2,0} e^{-i \omega t - \gamma t}.
\label{Eq:PerturbationAnsatz_Lin2}
\end{eqnarray}
Substituting Eqs.~(\ref{Eq:PerturbationAnsatz_Lin1}) and (\ref{Eq:PerturbationAnsatz_Lin2}) in Eqs.~(\ref{Eq:EOMPerturbation_Lin1}) and (\ref{Eq:EOMPerturbation_Lin2}), we obtain
\begin{equation}
\left(
\begin{array}{cc}
\left( \omega - i \gamma \right) - \left( \omega_{1} - i \kappa_{1} \right) & -g \\
-g & \left( \omega - i \gamma \right) - \left( \omega_{2} - i \kappa_{2} \right)
\end{array}
\right)
\left(
\begin{array}{cc}
\delta_{1,0} \\
\delta_{2,0}
\end{array}
\right) =
\left(
\begin{array}{cc}
0 \\
0
\end{array}
\right).
\end{equation}
Nontrivial solutions to this equation can exist if the determinant of the $2 \times 2$ matrix vanishes:
\begin{equation}
\left[ \left( \omega - i \gamma \right) - \left( \omega_{1} - i \kappa_{1} \right) \right] \left[ \left( \omega - i \gamma \right) - \left( \omega_{2} - i \kappa_{2} \right) \right] - g^2 = 0.
\end{equation}
The formula for the quadratic equation solution then gives
\begin{equation}
\left( \omega - \gamma \right) = \frac{\left( \omega_{1} - i \kappa_{1} \right) + \left( \omega_{2} - i \kappa_{2} \right)}{2} \pm \sqrt{\left[ \frac{\left( \omega_{1} - i \kappa_{1} \right) + \left( \omega_{2} - i \kappa_{2} \right)}{2} \right]^2 + g^2 - \left( \omega_{1} - i \kappa_{1} \right) \left( \omega_{2} - i \kappa_{2} \right)}.
\end{equation}

We can distinguish between level repulsion and level-attraction spectra by considering what happens when $\omega_1\approx\omega_2$ and $\kappa$ negligibly small. In this case, we obtain
\begin{equation}
\left( \omega - \gamma \right) = \frac{\omega_{1} + \omega_{2}}{2} \pm \sqrt{\left[ \frac{\omega_{1} - \omega_{2}}{2} \right]^2 + g^2}.
\end{equation}
When $g$ is real, we obtain level repulsion, while imaginary $g$ leads to level attraction. In the latter case, when $2|g|>\left| \omega_{1} - \omega_{2} \right|$, the two normal mode frequencies are equal.

\section{Appearance of multiple frequencies in oscillation modes}

To provide a simple picture explaining the emergence of oscillation modes that contain multiple frequencies, here we analyze a simpler dynamical system that exhibits a similar effect.

We consider the equation of motion
\begin{equation}
i \frac{dx}{dt} = \left( \omega_a - i \kappa \right) x + g x^* + \tau e^{-i \omega_d t}.
\label{Eq:SingleVariableEOM}
\end{equation}

The steady state solution cannot be given by a single-frequency function ($e^{-i\omega t}$), because the $x^*$ term would oscillate at the same frequency but with the opposite sign. Instead, we can look for a steady-state solution of the form
\begin{equation}
x = \sum_j A_j e^{-i \omega_j t}.
\end{equation}
We need only two terms to ensure that Eq.~(\ref{Eq:SingleVariableEOM}) is satisfied:
\begin{equation}
x = A_+ e^{-i \omega_d t} + A_- e^{i \omega_d t}.
\end{equation}
We then obtain the system of equations
\begin{eqnarray}
\left( \omega_d - \omega_a + i \kappa \right) A_+ - g A_-^* & = & \tau,
\\
\left( - \omega_d - \omega_a + i \kappa \right) A_- - g A_+^* & = & 0.
\end{eqnarray}
Straightforward algebraic manipulation then gives
\begin{equation}
\left( \omega_d - \omega_a + i \kappa + \frac{|g|^2}{\omega_d + \omega_a - i \kappa} \right) A_+ = \tau.
\end{equation}
This equation gives the steady-state oscillation amplitude $A_+$ as a function of driving strength $\tau$. The other amplitude in the steady-state solution ($A_-$) can then be evaluated straightforwardly. It is worth noting that the term $|g|^2/\left(\omega_d + \omega_a - i \kappa \right)$ contains the sum of two frequencies and will therefore generally contribute only a small correction to $A_+$ compared to the case $g=0$.

If we consider the normal modes of small perturbations around the steady state, or even the normal modes in the absence of a driving field ($\tau=0$), the solution is described by
\begin{equation}
x = \delta_+ e^{-i \omega t - \gamma t} + \delta_- e^{i \omega t - \gamma t}.
\end{equation}
A similar treatment as above gives the equation
\begin{equation}
\left( \omega - \omega_a + i \kappa - i \gamma + \frac{|g|^2}{\omega + \omega_a - \kappa + i \gamma} \right) \delta_+ = 0.
\end{equation}
Nontrivial solutions of this equation (with $\delta_+\neq 0$) can exist if
\begin{equation}
\left( \omega - \omega_a + i \kappa - i \gamma \right) \left( \omega + \omega_a - i \kappa + i \gamma \right) + |g|^2 = 0.
\end{equation}
Since we assume that $\omega_a\neq 0$, the product in the first term is real only if $\gamma=\kappa$. We then obtain the equation
\begin{equation}
\omega^2 = \omega_a^2 - |g|^2.
\end{equation}
A few point to note in this result are: (1) The oscillation frequency is smaller than $\omega_a$. (2) The solutions described by the above formula exist only when $|g|<\omega_a$. When $|g|>\omega_a$, the dynamical system becomes unstable and $x$ diverges to $\infty$. (3) The normal mode described by these equations contains two terms with the oscillation frequencies $\pm\omega$. In this case, the frequencies in the two terms are equal. In the two-mode problem described in Secs.~\ref{Sec:Optical} and \ref{Sec:Acoustic}, we obtain a sum of terms with different frequencies.

\end{document}